\documentclass{jkas}

\def\year{2025} 
\def\volume{58} 
\def\issue{2} 
\def\beginpage{1} 
\def\received{May 13, 2025} 
\def\accepted{June 27, 2025} 
\def\published{January 1, 2023} 
\date{Received \received; Accepted \accepted; Published \published}

\hypersetup{colorlinks=true,linkcolor=blue,citecolor=blue,filecolor=cyan,urlcolor=magenta}

\def\arcsec{\hbox{$^{\prime\prime}$}}

\def\smallplot#1{\centering \leavevmode
\includegraphics[width=89.5mm]{#1} }

\def\smallplottwo#1#2{\centering \leavevmode
\includegraphics[width=89.5mm]{#1}
 \hfil \includegraphics[width=89.5mm]{#2} }

\def\smallplotfour#1#2#3#4{\centering \leavevmode
\includegraphics[width=89.5mm]{#1} \includegraphics[width=89.5mm]{#2}
\hfil \includegraphics[width=89.5mm]{#3} \includegraphics[width=89.5mm]{#4}}

\def\smallploteight#1#2#3#4#5#6#7#8{\centering \leavevmode
\includegraphics[width=89.5mm]{#1} \includegraphics[width=89.5mm]{#2}
\hfil \includegraphics[width=89.5mm]{#3} \includegraphics[width=89.5mm]{#4}
\hfil \includegraphics[width=89.5mm]{#5} \includegraphics[width=89.5mm]{#6}
\hfil \includegraphics[width=89.5mm]{#7} \includegraphics[width=89.5mm]{#8}}

\def\largeplottwo#1#2{\centering \leavevmode
\includegraphics[width=135mm]{#1}
 \hfil \includegraphics[width=135mm]{#2} }

\title{%
Infrared Variability of Carbon Stars in the LMC Observed with NEOWISE-R
}

\author[$\star$]{Kyung-Won Suh}{0000-0001-9104-9763}

\affil[1]{Department of Astronomy and Space Science, Chungbuk National University,
Chungcheongbuk-do 28644, Republic of Korea}

\def\corrauthor{%
K.-W. Suh, \email{kwsuh@chungbuk.ac.kr}
}

\def\runningauthor{%
Suh
}

\def\runningtitle{%
Infrared Variability of Carbon Stars in the LMC Observed with NEOWISE-R
}

\def\keywords{%
stars: AGB and post-AGB --- stars: oscillations --- circumstellar matter
--- dust, extinction --- infrared: stars --- radiative transfer
}

\def\abstracttext{
We investigate the infrared variability of carbon stars in the Large Magellanic
Cloud (LMC). Our sample consists of 11,134 carbon stars identified in both visual
and infrared bands. Among these, 1,184 objects are known Mira variables based on
the Optical Gravitational Lensing Experiment III (OGLE-III) observations. We study
the infrared variability of the entire sample using the Wide-field Infrared
Survey Explorer (WISE) photometric data spanning the past 16 years, including the
AllWISE multiepoch data and the Near-Earth Object WISE Reactivation (NEOWISE-R)
2024 final data release. We generate light curves using WISE observations in the
W1 and W2 bands and compute Lomb-Scargle periodograms for all sample stars. From
the WISE light curves, we derive reliable variability parameters for 1,615
objects. Among these, we identify 672 objects exhibiting clear Mira-like
variations: 445 of these are previously known Miras from OGLE-III, while 227 are
candidates for new Mira variables identified from the WISE data. We establish
period-magnitude and period-color relations in both visual and infrared bands for
the known Miras and the newly identified candidates from WISE data. We anticipate that
these relations will serve as valuable references for studying carbon stars in
other galaxies, including the Milky Way.
}

\begin{document}
\jkashead 


\begin{table*}
\small
\setlength{\tabcolsep}{4pt}
\centering
\caption{Sample of Carbon stars in the LMC and our Galaxy\label{tab:tab1}}
\begin{tabular}{llllllll}
\hline
\hline
Group  &Reference &Total Number & Selected & WISE$^a$ & WISE\_P$^a$ & WISE\_S$^a$  \\
\hline
\hline
CAGB-LMC-SAGE & \citet{Riebel2012} & 7,308$^b$ & 7,308  & 1,532 (664 [212]) & 1,198 (605 [206]) & 334 (47 [6])  \\
CAGB-LMC-SAGE-S$^c$ & SJ$^d$ & 151 & 151  &  &  &   \\
CS-LMC-K & \citet{kontizas2001} & 7,760 & 3,826$^e$  & 83 (20 [15])  & 53 (16 [12]) & 30 (4 [3])  \\
\hline
CAGB-LMC-W$^f$ &   &  &   & 1,615 (672 [227]) & 1,251 (621 [218]) & 364 (51 [9])  \\
\hline
LPV-OGLE-III-LMC & \citet{sus09} & 91,995 & 1,184$^g$  &  &  &   \\
\hline
CAGB-MW & \citet{suh2024} & 7,163  & 7,163 &   &  &  \\
\hline
\end{tabular}
\begin{flushleft}
$^a$The number of objects for which meaningful variability parameters were derived from the WISE light curves
(see Section~\ref{sec:neo-a} for details).
WISE\_P designates objects for which primary periods were selected, whereas
WISE\_S refers to those where secondary or tertiary periods were adopted. Values
in parentheses indicate the subset exhibiting strong Mira-like variability ($R^2
> 0.8$), and values in brackets denote a further subset representing candidates
for newly identified Mira variables from WISE data.
$^b$Newly identified CAGB objects are added; 3,933 objects are duplicates of CS-LMC-K objects.
$^c$Subgroup of CAGB-LMC-SAGE; objects Identified from the SAGE IRS spectroscopy.
$^d$\citet{sloan2016} and \citet{jones2017}.
$^e$One object, KDMK 1480, is excluded because it is not a real object and duplicately identified as a carbon star KDMK 1483 by SIMBAD;
and 3,933 objects are excluded because they are duplicates of CAGB-LMC-SAGE objects.
$^f$Combined objects from the CAGB-LMC-SAGE and CS-LMC-K samples with meaningful variability parameters derived from WISE light curves.
$^g$Mira variables identified as CAGB stars, originating from either the CAGB-LMC-SAGE or CS-LMC-K samples.
\end{flushleft}
\end{table*}

\section{Introduction} \label{sec:intro}

Carbon-rich asymptotic giant branch (CAGB) stars are widely considered to be the
evolutionary descendants of M-type oxygen-rich AGB (OAGB) stars in the early
stages of the AGB phase (\citealt{iben1983}). Intermediate-mass OAGB stars (with
masses between 1.55 and 4 $M_{\odot}$ at solar metallicity) can undergo third
dredge-up events triggered by thermal pulses during the AGB phase. These
dredge-ups can raise the carbon-to-oxygen (C/O) ratio above unity, halting the
production of oxygen-rich dust and marking the transition to visually observable
CAGB stars (\citealt{groenewegen1995}). Following this transition, carbon-rich
dust begins to form, and the stars evolve further into infrared (IR) carbon
stars, which are CAGB stars surrounded by thick carbon-rich dust shells and
exhibiting elevated mass-loss rates (e.g., \citealt{suh2000}).

Nearly all AGB stars are long-period variables (LPVs), characterized by
large-amplitude pulsations (e.g., \citealt{hofner2018}). It is generally thought
that a giant star evolves from a small-amplitude red giant (SARG) to a
semiregular variable (SRV), and eventually becomes a Mira variable. This
progression involves a reduction in the number of pulsation modes and an increase
in both pulsation period and amplitude (\citealt{swu13}). While many SRVs and a
significant fraction of SARGs may also be in the AGB phase, Mira variables are
almost certainly AGB stars. They typically pulsate in the fundamental mode and
occupy a distinct sequence in the period-luminosity diagram (e.g.,
\citealt{swu13}).

While most carbon stars are CAGB stars, the category also includes other types.
Typically, classical (also referred to as intrinsic or type C-N) carbon stars are
understood to be CAGB stars. In contrast, there are non-classical, or extrinsic,
carbon stars that are not in the AGB evolutionary stage. These include Barium
(Ba) stars, CH (or C-H) stars, dwarf carbon (dC) stars, J-type (C-J) stars, and
early R-type (R-hot or RH) stars (e.g., \citealt{green2013}; \citealt{suh2024}).

Thanks to the Optical Gravitational Lensing Experiment III (OGLE-III) project
(\citealt{sus09}) and Spitzer Space Telescope Legacy program `Surveying the
Agents of a Galaxy Evolution' (SAGE; \citealt{meixner2006}), a much large number
of carbon stars in the LMC are identified and studied (e.g., \citealt{suh2020}).

In 2009, the Wide-field Infrared Survey Explorer (WISE; \citealt{wright2010})
started mapping the sky. The AllWISE multiepoch photometry table obtained in
2009-2010 provided the photometric data in four bands (3.4, 4.6, 12, and 22
$\mu$m; W1, W2, W3, and W4). And the Near-Earth Object WISE Reactivation
(NEOWISE-R) mission (\citealt{mainzer2014}) provided photometric data in W1 and
W2 bands for last 10 years.

In this work, we investigate variability of carbon stars in W1[3.4] and W2[4.6]
bands over the past 16 years, utilizing AllWISE multi-epoch photometry from
2009–2010 and NEOWISE-R data from the 2024 final data release. The NEOWISE-R data
dataset includes 21 epochs, with two observations per year from 2014 to 2023 and
one additional observation in 2024.

In Section~\ref{sec:sample}, we provide comprehensive lists of known carbon stars
in the LMC. In Section~\ref{sec:neo}, we investigate variability of the carbon
stars using the WISE data over the past 16 years. In Section~\ref{sec:pul}, we
establish period-color and period-magnitude relations in both visual and IR bands
for the known Miras and candidates for newly identified Miras from WISE.

In Section~\ref{sec:models}, we describe the theoretical radiative transfer
models used for the dust shells surrounding CAGB stars. Section~\ref{sec:cmd}
presents various color-magnitude diagrams (CMDs) and two-color diagrams (2CDs)
for carbon stars and Mira variables, juxtaposed with theoretical models. Finally,
Section~\ref{sec:sum} consolidates and summarizes the key findings of this study.

\section{Sample Stars\label{sec:sample}}

We use catalogs of carbon stars in the LMC from the available literature.
Table~\ref{tab:tab1} lists the reference, total number of objects, and numbers of
objects for which meaningful variability parameters were derived from the WISE
light curves.

\subsection{Carbon stars in the LMC\label{sec:magb}}

Analyzing the Spitzer data of the SAGE program, \citet{Riebel2012} presented a
list of 7,293 CAGB objects. The Spitzer Infrared Spectrograph (IRS; $\lambda$ =
5.2-38 $\mu$m) has taken high resolution spectra for many AGB stars in the LMC.
When we compile the lists from \citet{sloan2016} and \citet{jones2017} there are
151 CAGB stars in the LMC which are identified from the SAGE IRS data. These SAGE
IRS sample stars (CAGB-LMC-SAGE-S objects) that are identified by the IRS spectra
would be more reliable sample of CAGB stars in the LMC.

When we compare these 151 CAGB-LMC-SAGE-S objects with the list of
\citet{Riebel2012}, we found that the total number of the CAGB stars in the LMC,
identified from the SAGE program, is 7,308 (CAGB-LMC-SAGE objects; see
Table~\ref{tab:tab1}).  Note that all of the objects in the CAGB-LMC-SAGE-S
objects are already included in the CAGB-LMC-SAGE sample.

\citet{kontizas2001} presented a catalog of 7,760 carbon stars in the LMC
(CS-LMC-K objects; see Table~\ref{tab:tab1}). Most of them are CAGB stars and
3,933 objects are duplicates of CAGB-LMC-SAGE objects. Some of these objects
could be extrinsic carbon stars. Table~\ref{tab:tab1} lists the sample carbon
stars in the LMC.

OGLE-III presented 91,995 LPVs in the LMC (\citealt{sus09}). The sample is
composed of 1,663 Mira variables, 11,132 SRVs, and 79,200 SARGs. There are 1,184
Mira variables from OGLE-III in the list of carbon stars in the LMC
(CAGB-LMC-SAGE and CS-LMC-K objects).

\begin{table}
\footnotesize
\setlength{\tabcolsep}{4pt}
\caption{Photometric bands and zero magnitude flux values \label{tab:tab2}}
\centering
\begin{tabular}{lllll}
\hline \hline
Band &$\lambda_{ref}$ ($\mu$m)	&ZMF (Jy) &Telescope &Reference$^1$ 	\\
\hline
G[0.6]  &0.622	&	3229	&	Gaia DR3    & \citet{rimoldini2023}	\\
Rp[0.8] &0.777	&	2555	&	Gaia DR3   & \citet{rimoldini2023}	\\
I[0.8]  &0.806	&	2400	&	OGLE-III   & \citet{sus09}	\\
J[1.2]  &1.235	&	1594	&	2MASS	& \citet{cohen2003}\\
K[2.2]  &2.159	&	666.7	&	2MASS	& \citet{cohen2003}\\
W1[3.4]	&3.35	&	306.682	&	WISE    & \citet{jarrett2011}	\\
W2[4.6]	&4.60	&	170.663	&	WISE    & \citet{jarrett2011}	\\
W3[12]$^a$	&12.0 	& 28.3	&	WISE    & \citet{jarrett2011}	\\
W4[22]	&22.08	&	8.284	&	WISE    & \citet{jarrett2011}	\\
\hline
\end{tabular}
\begin{flushleft}
$^a$For W3[12], $\lambda_{ref}$ and ZMF values are adopted from \citet{suh2020}.
\end{flushleft}
\end{table}

\subsection{Photometric Data and Cross-Matching\label{sec:photdata}}

Table~\ref{tab:tab2} lists the photometric bands used in this study, along with
their reference wavelengths ($\lambda_{ref}$) and zero-magnitude flux (ZMF)
values. These values are useful for comparing observed magnitudes with
theoretical model fluxes (see Section~\ref{sec:models}).

For the full sample of carbon stars in the LMC, we performed cross-identification 
with sources in the OGLE-III catalog by matching to the nearest counterparts 
within 3$\arcsec$. Similarly, cross-identification with sources in the 2MASS and 
WISE catalogs was carried out using the same 3$\arcsec$ matching radius. The mean 
matching radius for counterparts is 0.13$\arcsec$ $\pm$ 0.14 for OGLE-III, 
0.13$\arcsec$ $\pm$ 0.15 for 2MASS, and 0.16$\arcsec$ $\pm$ 0.18 for WISE.

\begin{figure*}
\centering
\largeplottwo{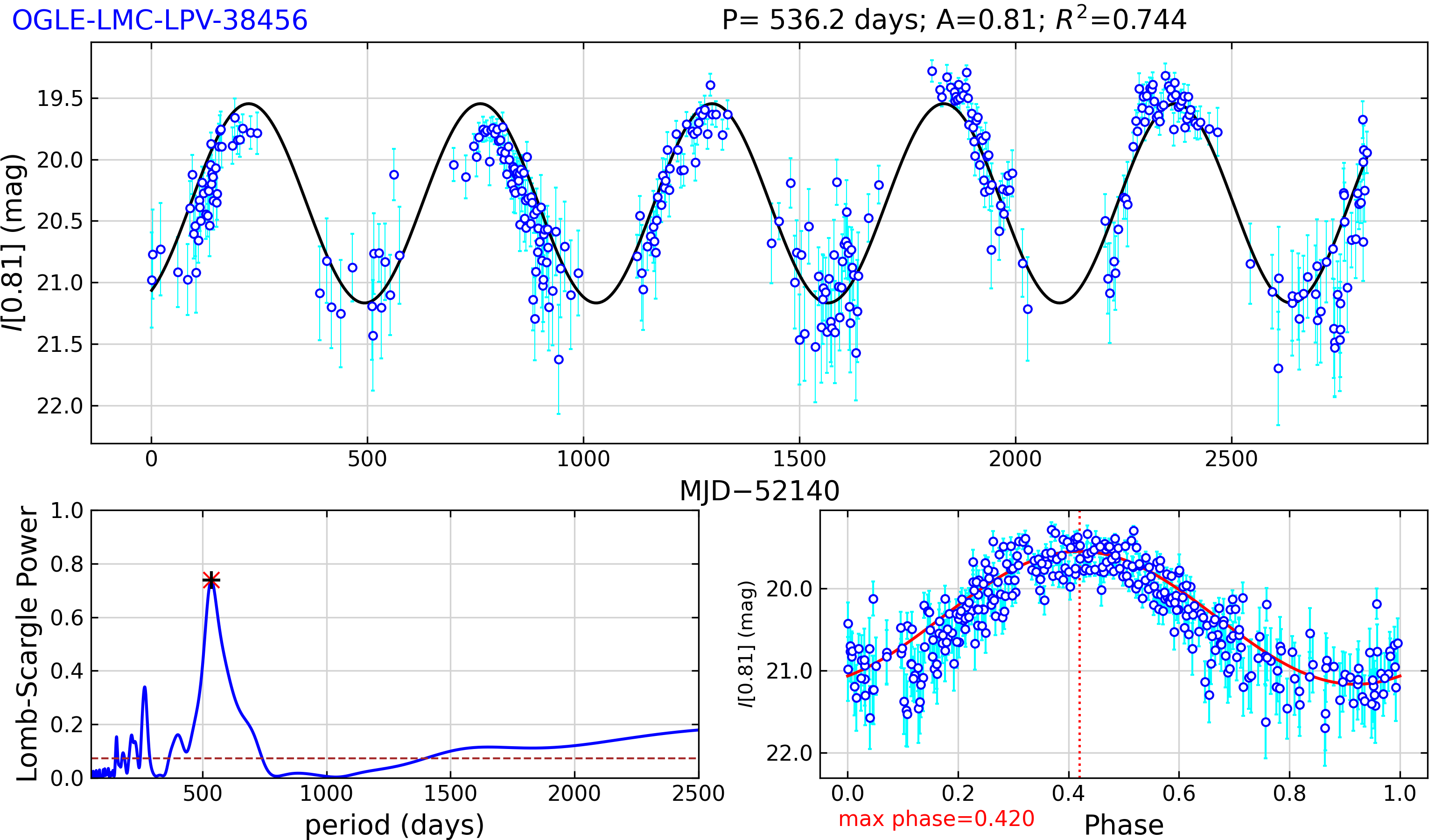}{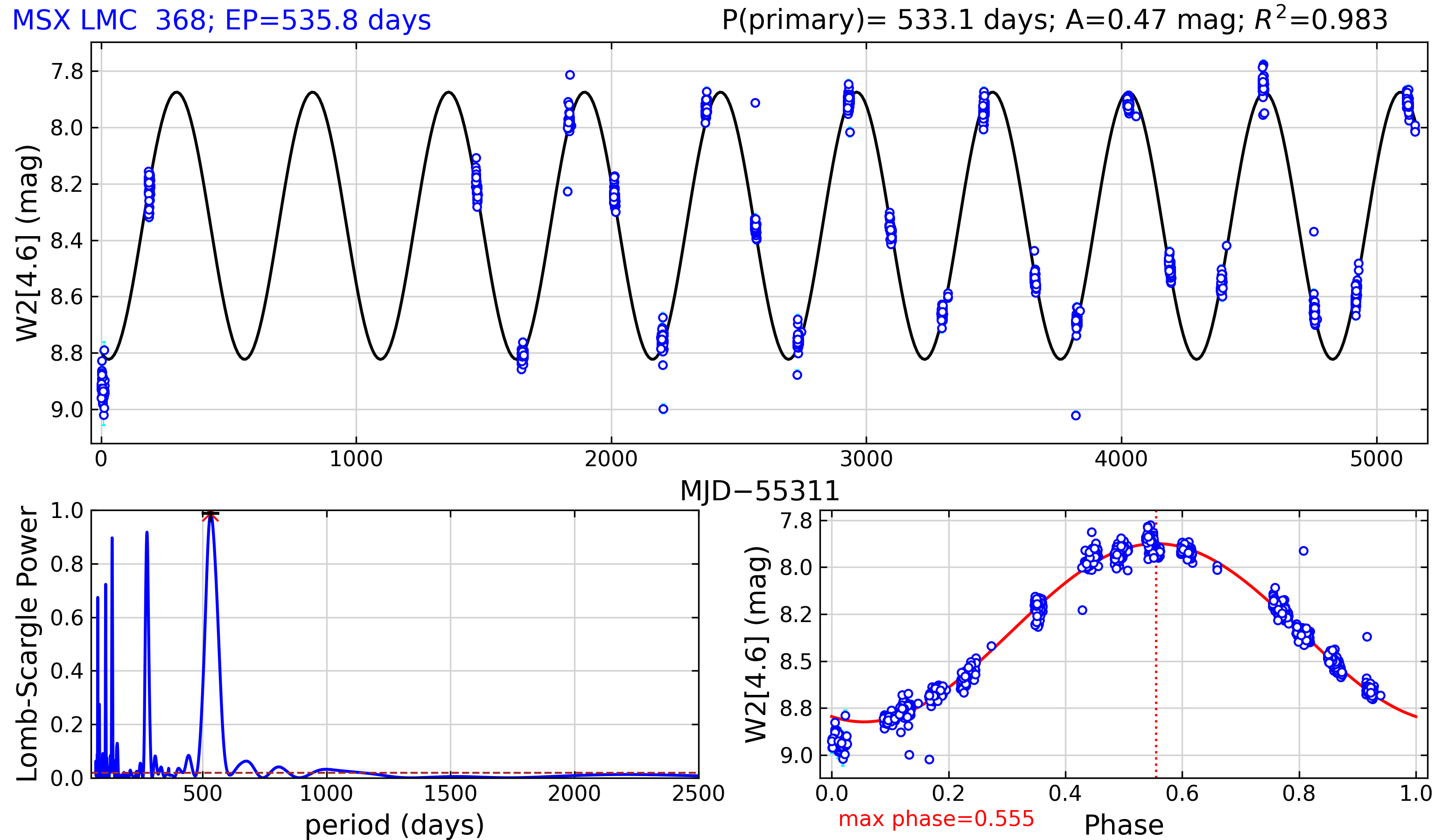}\caption{Lomb-Scargle periodograms based on light curves from
OGLE-III in the I[0.8] band (top panel) and WISE W2[4.6] band (bottom panel)
for a CAGB star MSX LMC 368 (OGLE-LMC-LPV-38456). See Section~\ref{sec:neo-a} for details.}
\label{f1}
\end{figure*}

\section{Finding IR variations of Carbon stars from WISE data\label{sec:neo}}

Mira variables are known to exhibit the strongest correlation between IR fluxes
and pulsation periods among pulsating variable stars (e.g., \citealt{sus09}).
Unlike Galactic Miras, those in the LMC provide a clearer view of the
period–magnitude relation (PMR) due to their relatively uniform distances.

\citet{suh2021} investigated the variability of Galactic AGB stars using WISE 
data. \citet{suh2024} further explored the variability of Galactic CAGB stars 
with a new sample stars and updated NEOWISE-R data.

To examine the variability of carbon stars in the LMC using WISE W1[3.4] and
W2[4.6] data over the past 16 years, we use multi-epoch photometry from AllWISE
(2009–2010) and the final NEOWISE-R data release (2024), which provides
observations from 21 epochs, with typically two per year between 2014 and 2023,
and one in 2024.

To detect Mira-like variability in the WISE light curves, we applied a simple 
sinusoidal model and analyzed the data using the Lomb–Scargle periodogram, a 
commonly used statistical method for identifying and characterizing periodic 
signals in unevenly spaced observations (e.g., \citealt{zechmeister2009}; 
\citealt{vanderPlas2018}). We focused on variations with periods longer than 50 
days and shorter than 2,500 days, consistent with known Mira variables. The 
Lomb–Scargle periodograms were computed using the 
AstroPy\footnote{\url{https://docs.astropy.org/en/stable/timeseries/lombscargle.html}} 
implementation (version 7.01), specifically employing the autopower option with 
the chi2 method, which leverages the equivalence between the periodogram value at 
each frequency and a least-squares sinusoidal fit to the data. This method also 
allows for extensions to multiple Fourier terms.

For each light curve, we identified the most significant peak in the periodogram 
and adopted the corresponding period, referred to as the best period, to fit a 
sinusoidal model with four parameters: period, amplitude, phase, and offset. To 
evaluate the quality of the fit, we calculated the signal-to-noise ratio (SNR), 
which measures the amplitude relative to the residual scatter, and the 
coefficient of determination ($R^2$), which indicates how well the sinusoidal 
model accounts for the observed variability.

In contrast to light curves from other surveys such as OGLE-III, the WISE data, 
collected at regular six-month intervals, often yield periodograms with multiple 
peaks of comparable power. This sampling pattern can make it difficult to 
identify accurate pulsation periods using WISE data alone (see 
\citealt{suh2021}). In some cases, the true period may correspond to a secondary 
or tertiary peak in the periodogram, while the primary peak may reflect an alias 
or harmonic, potentially obscuring the correct periodicity (e.g., 
\citealt{vanderPlas2018}). Selecting a secondary or tertiary peak results in 
lower values of both the SNR and $R^2$, as these peaks typically correspond to 
less accurate fits of the sinusoidal model to the observed data.

\begin{figure*}
\centering
\smallploteight{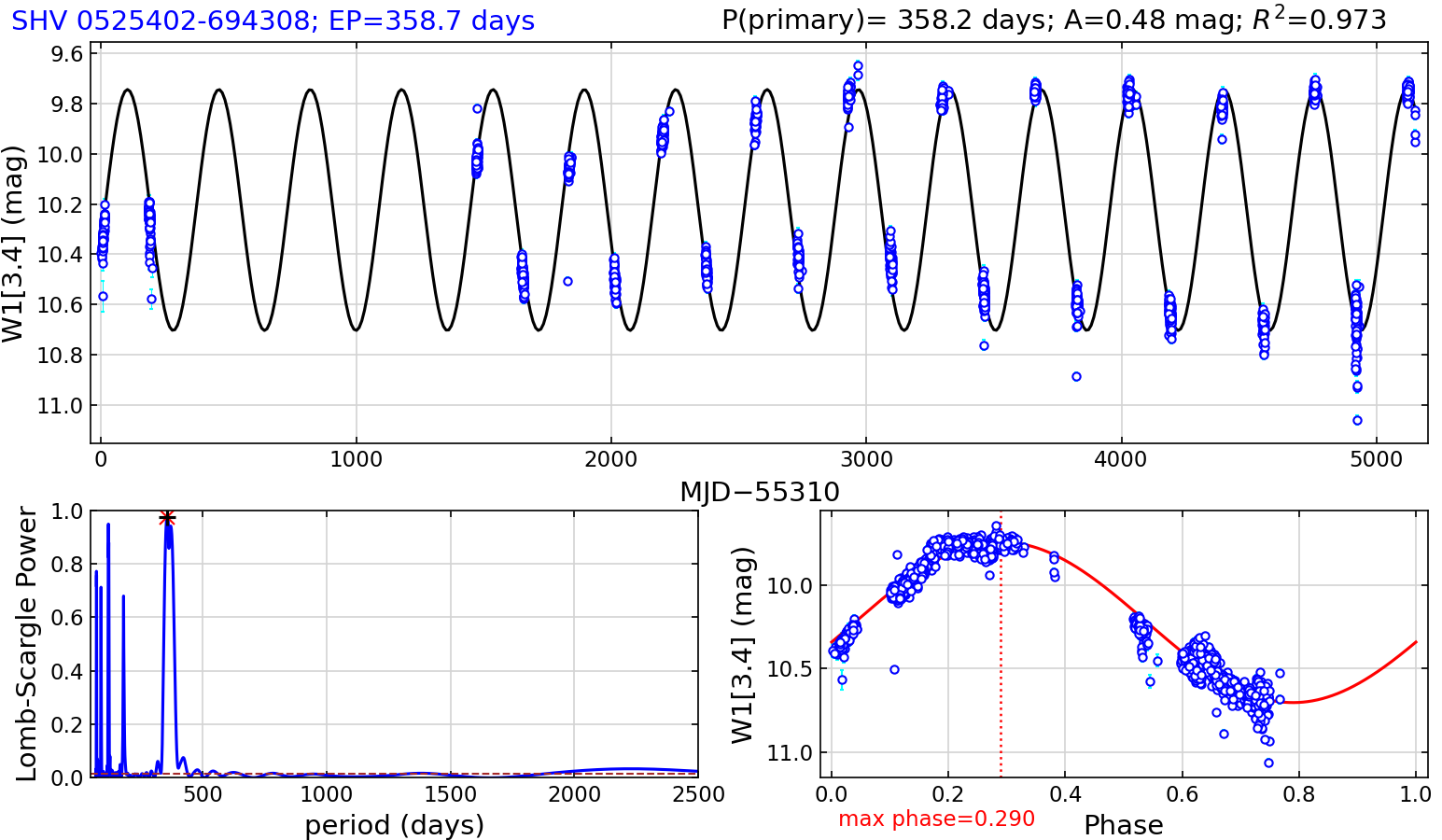}{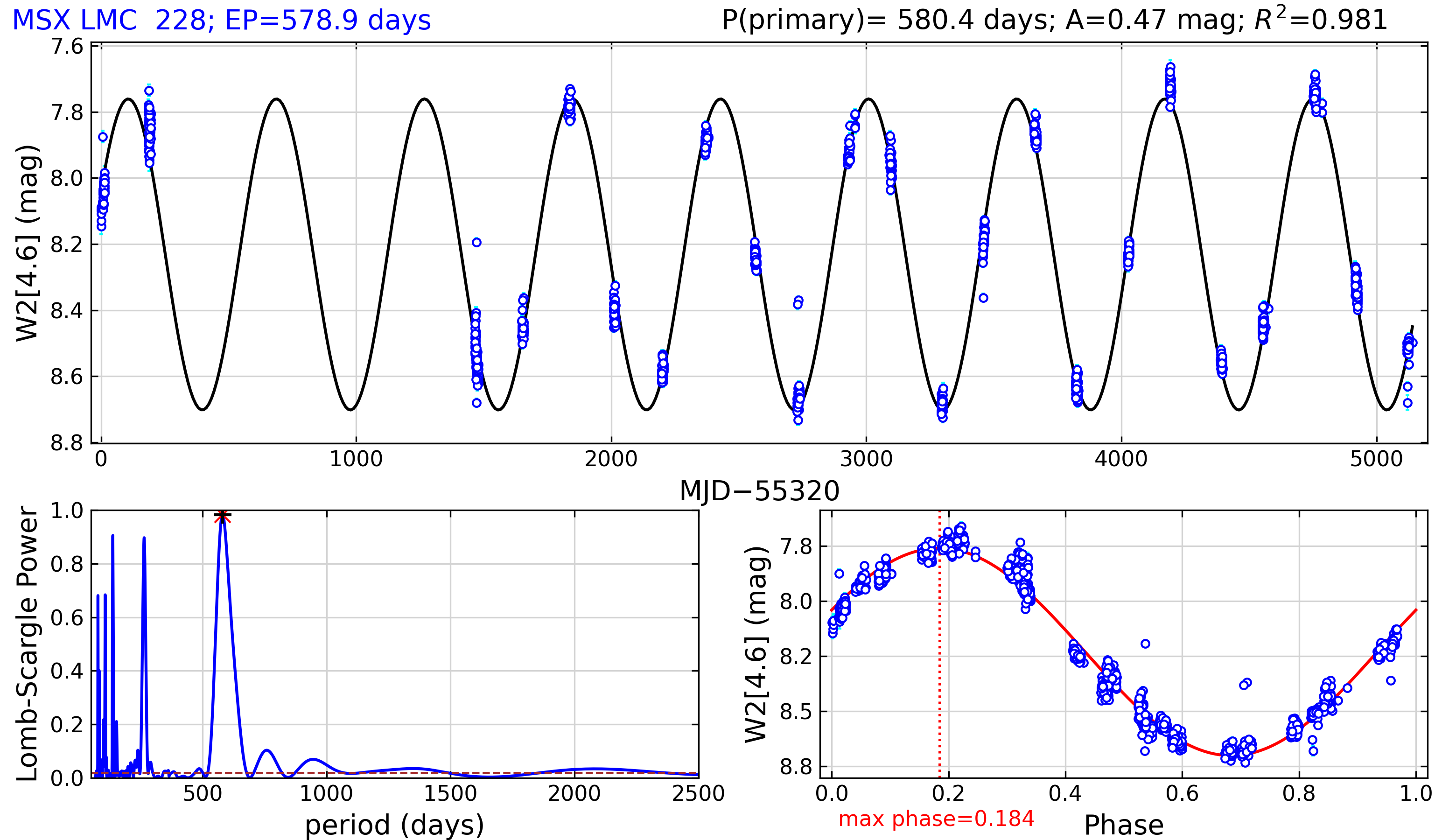}{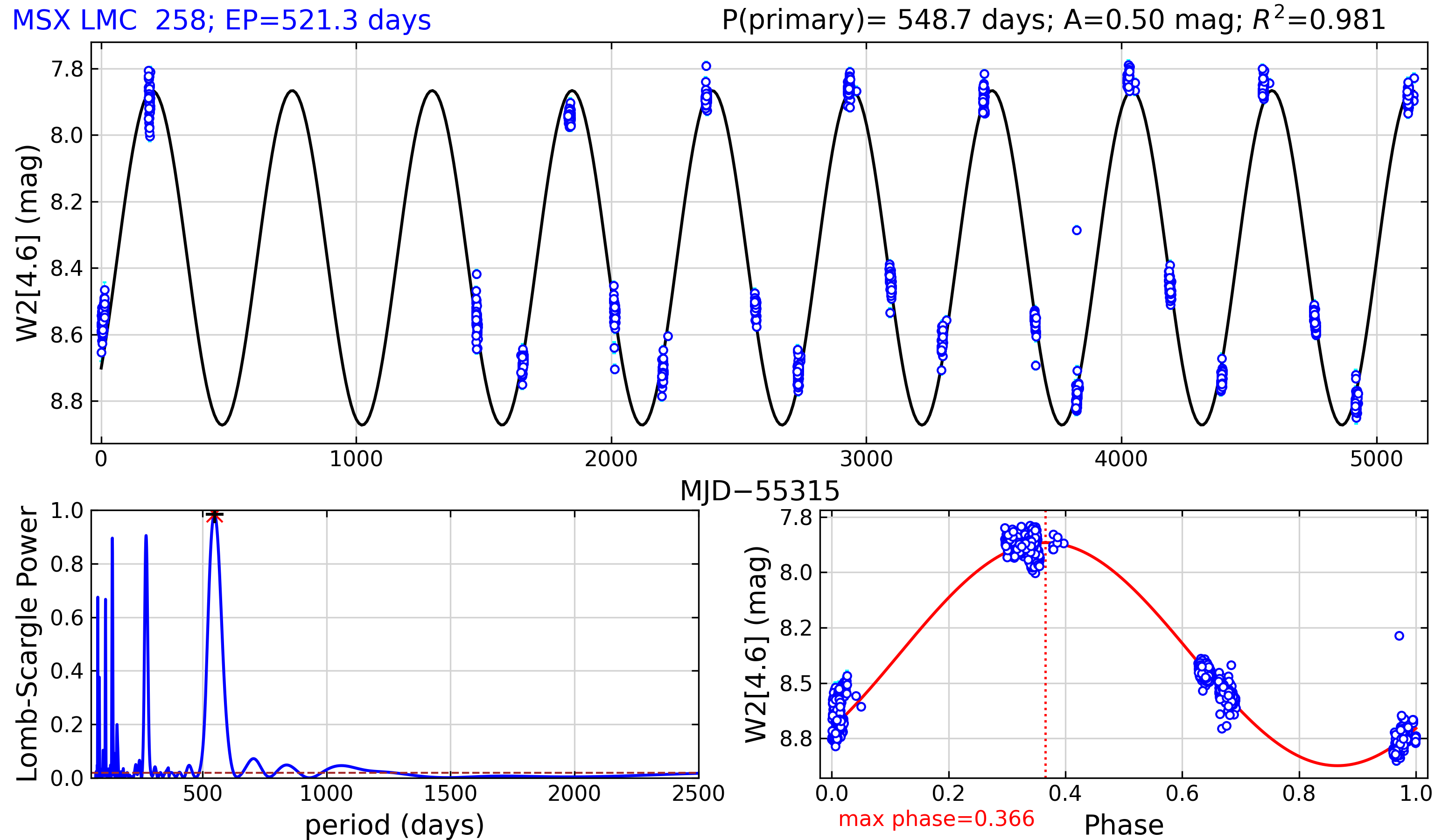}{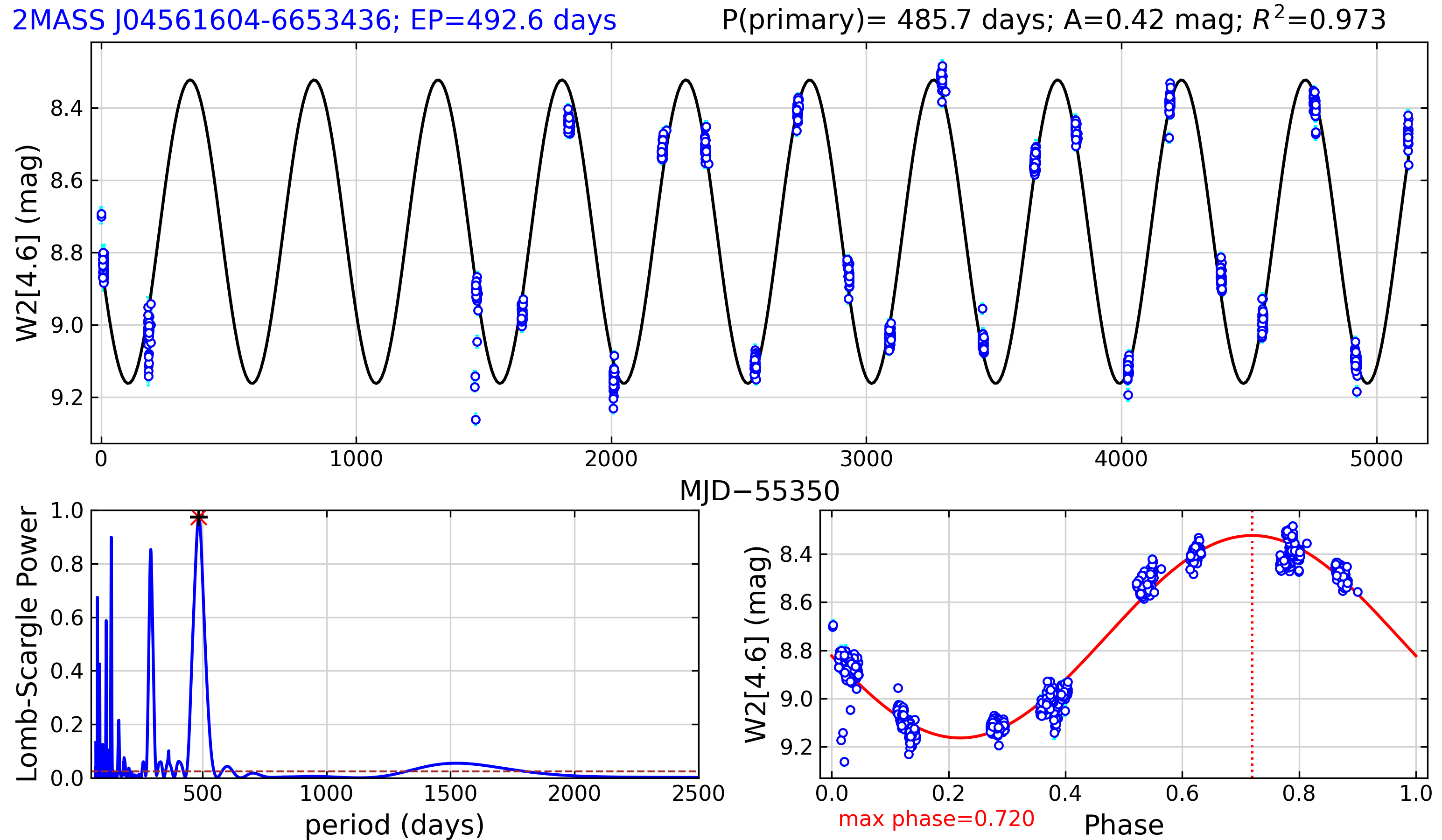}{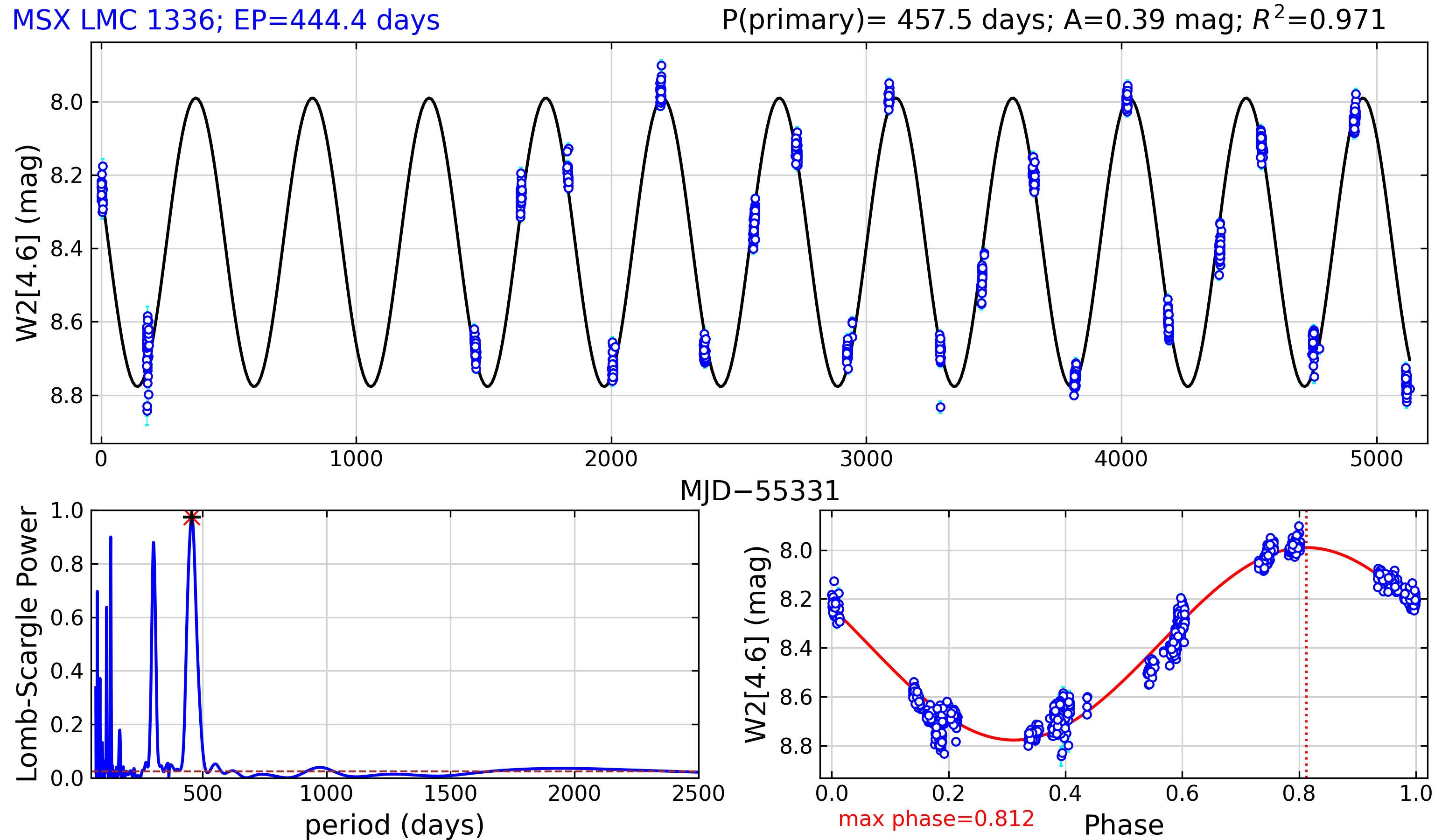}{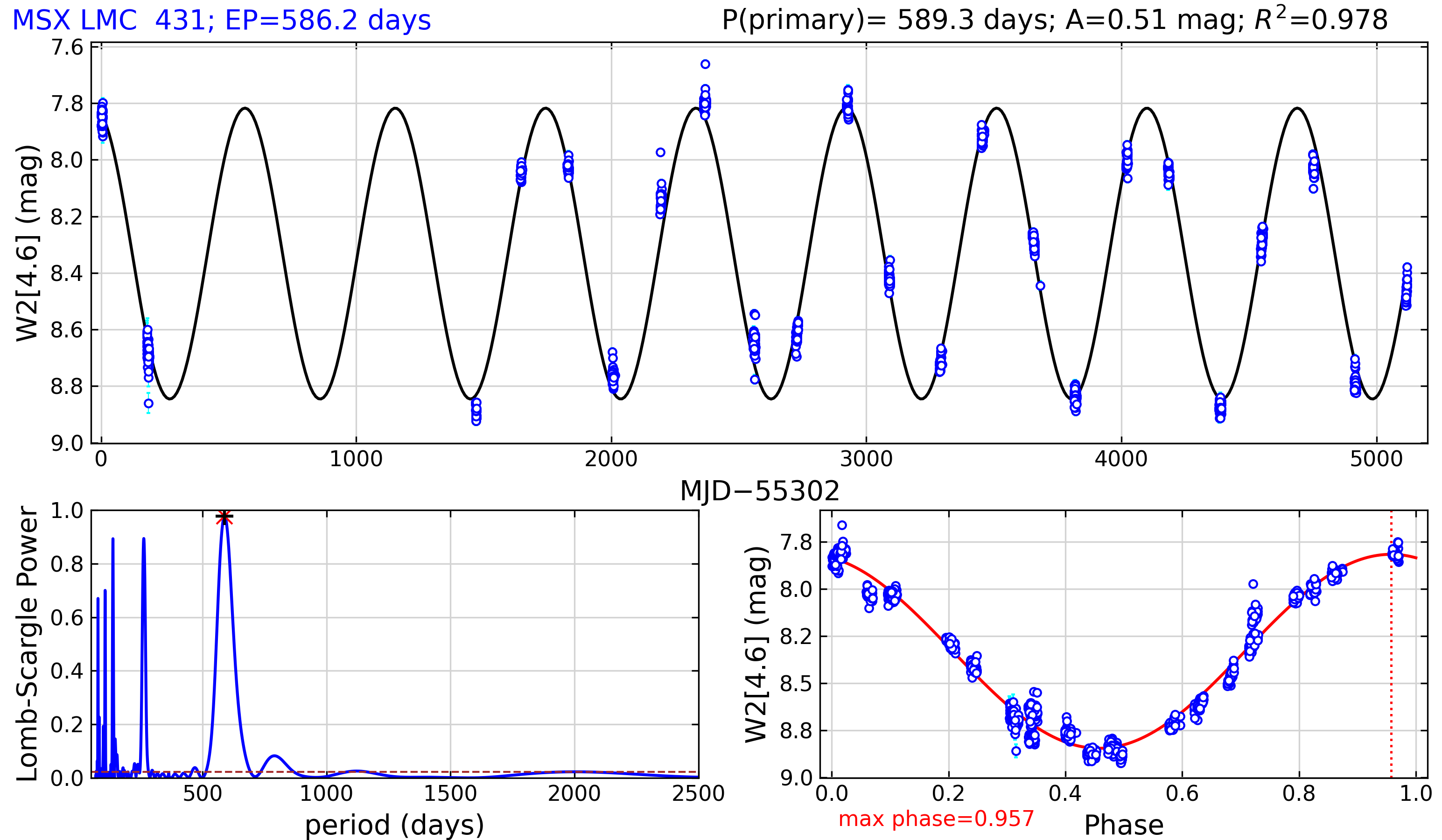}{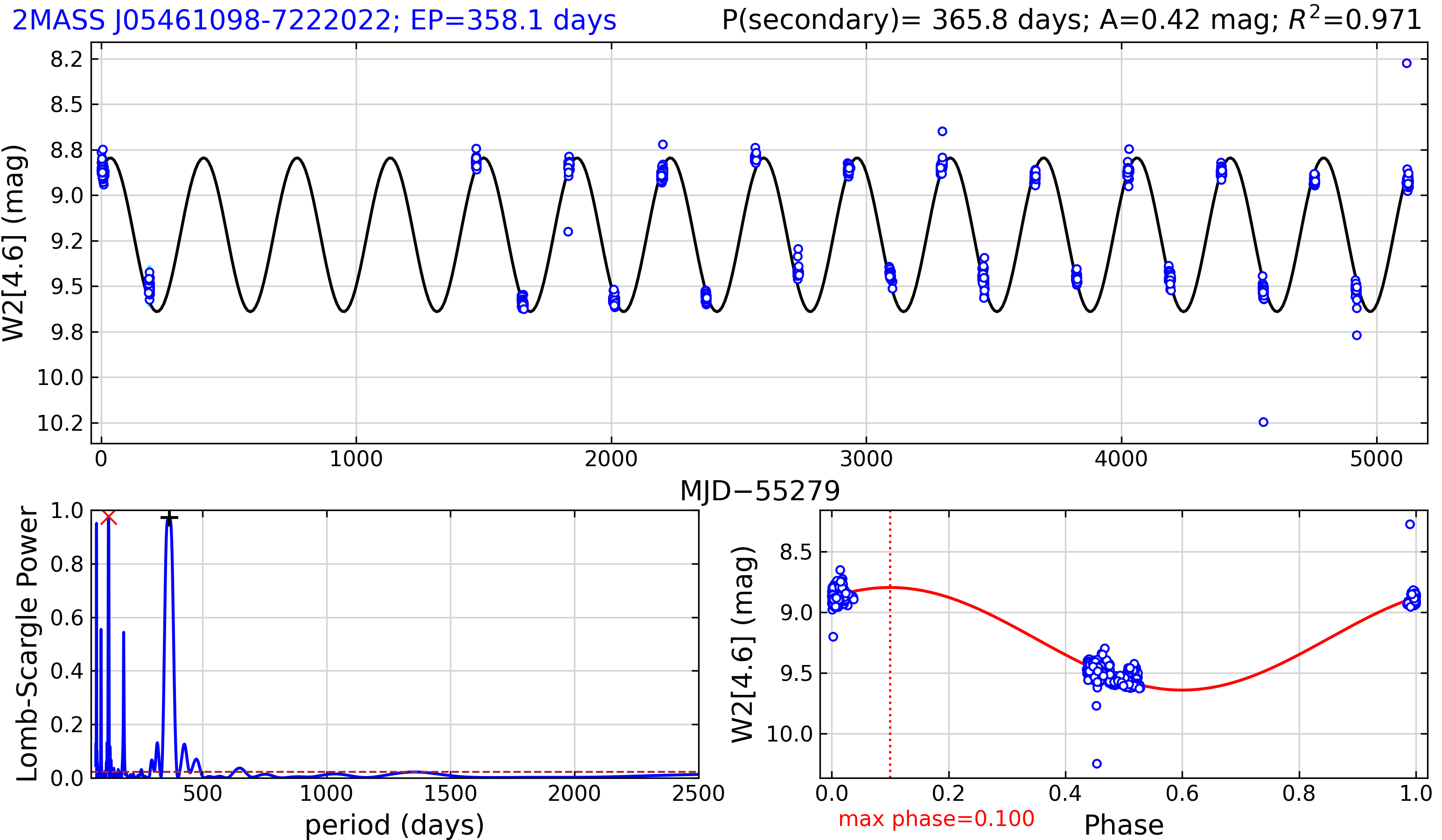}{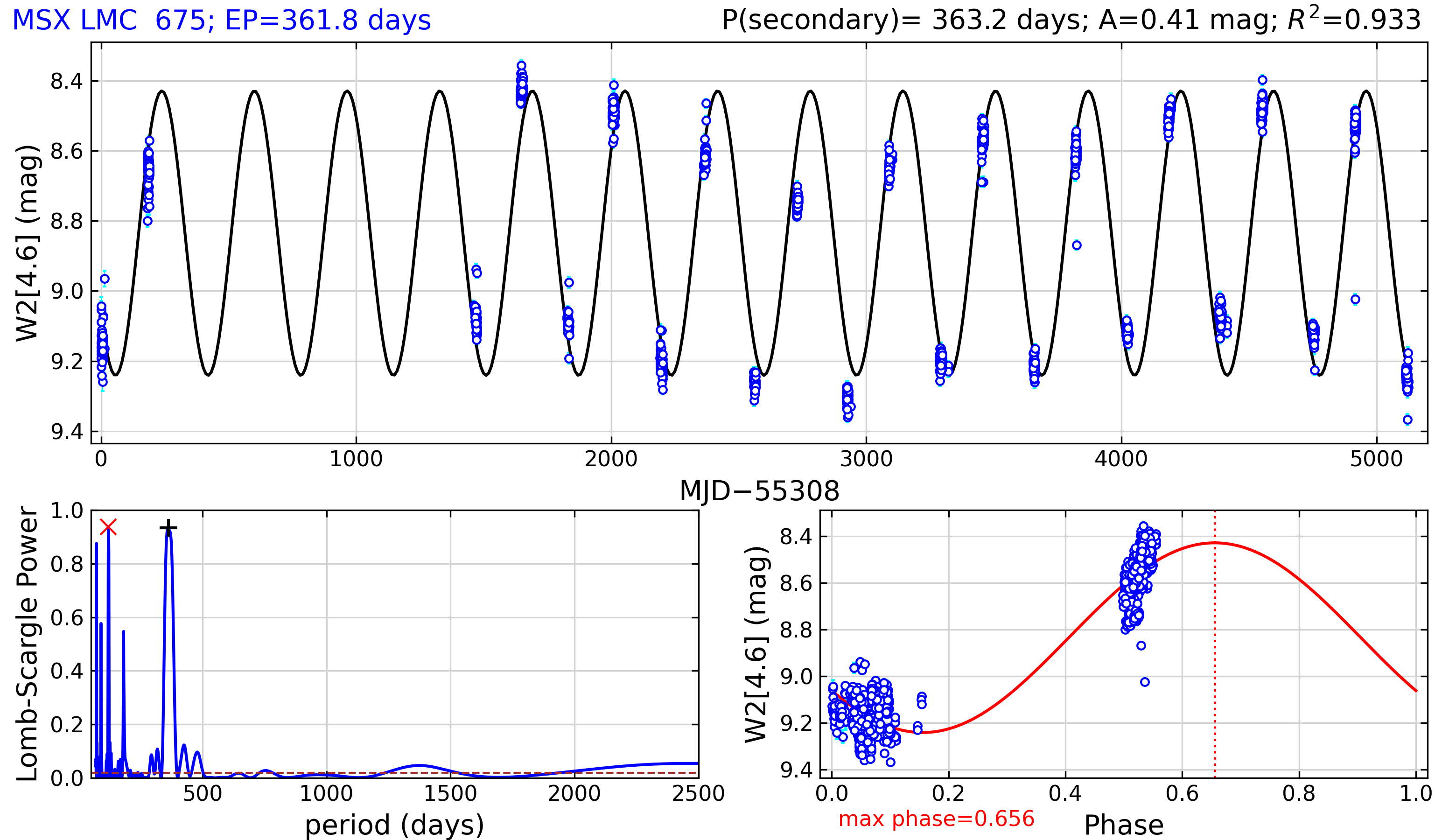}
\caption{Lomb-Scargle periodograms using WISE light curves for eight CAGB stars known as Miras according to OGLE-III.
The expected period (EP) denotes the period from OGLE-III (see Section~\ref{sec:neo-mc}).
Note that the secondary peak is used for the last two objects.
The periods obtained from WISE data are similar to those from OGLE-III.
See Section~\ref{sec:neo-a} for details.}
\label{f2}
\end{figure*}

\begin{figure*}
\centering
\smallploteight{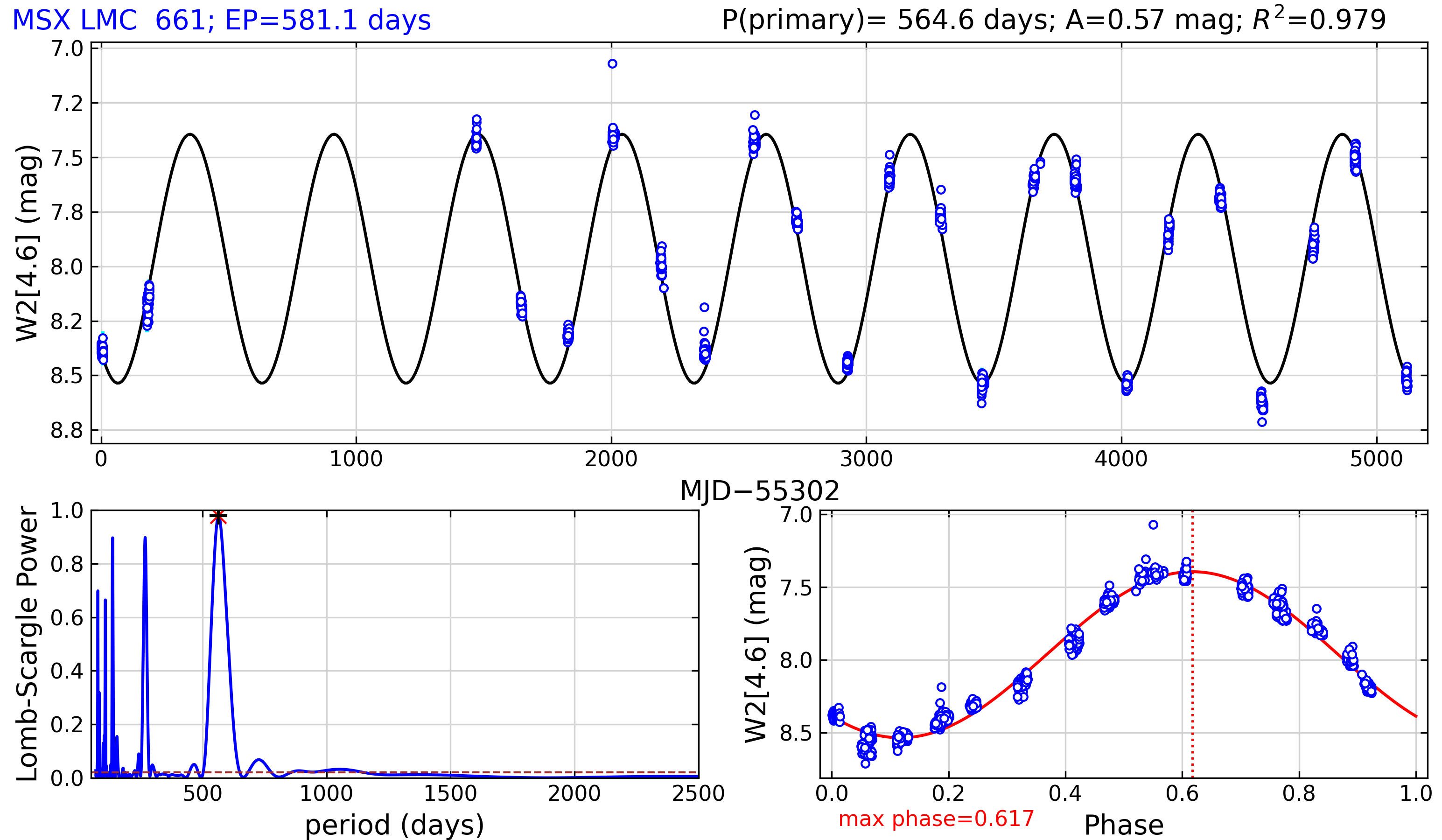}{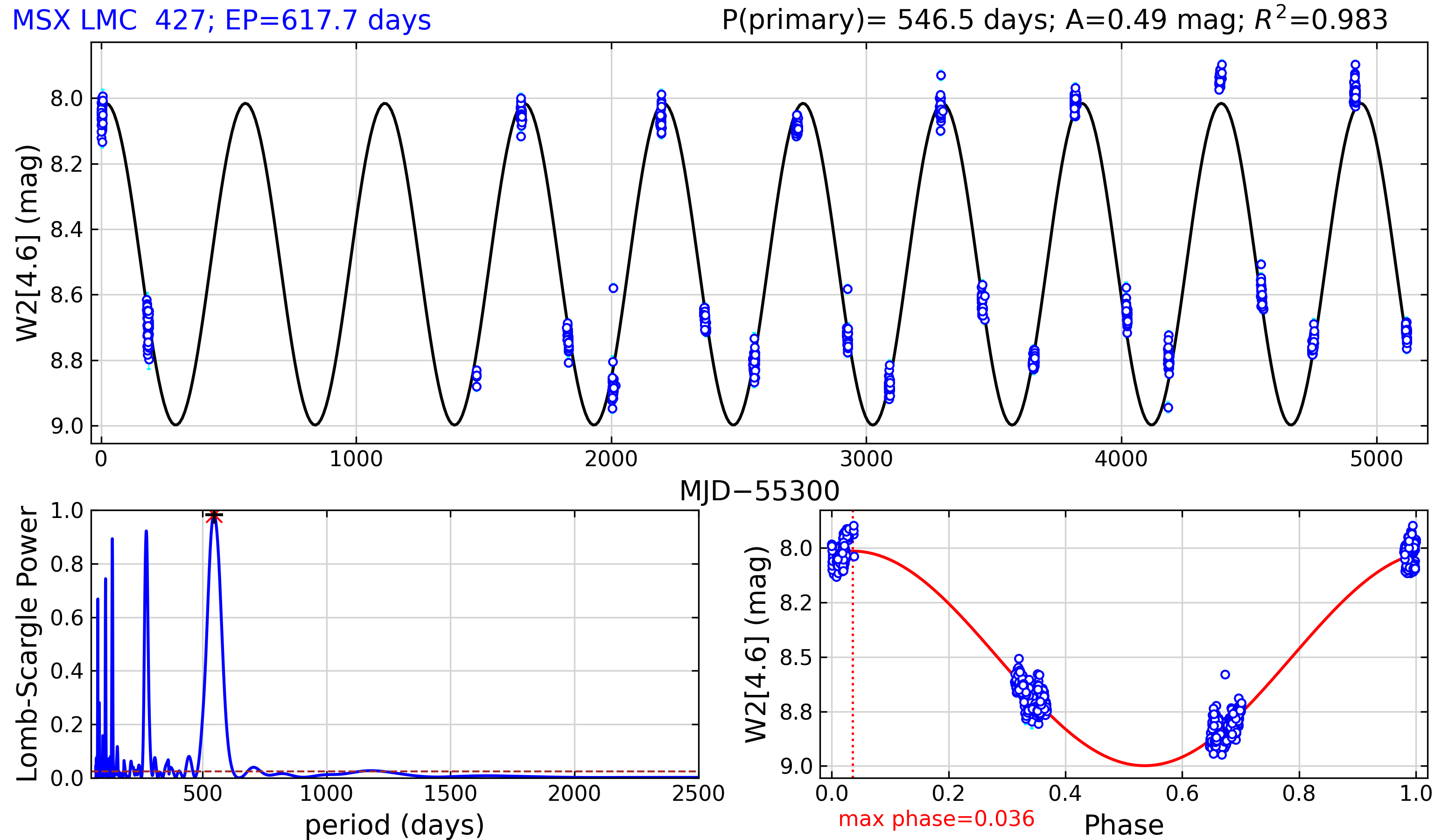}{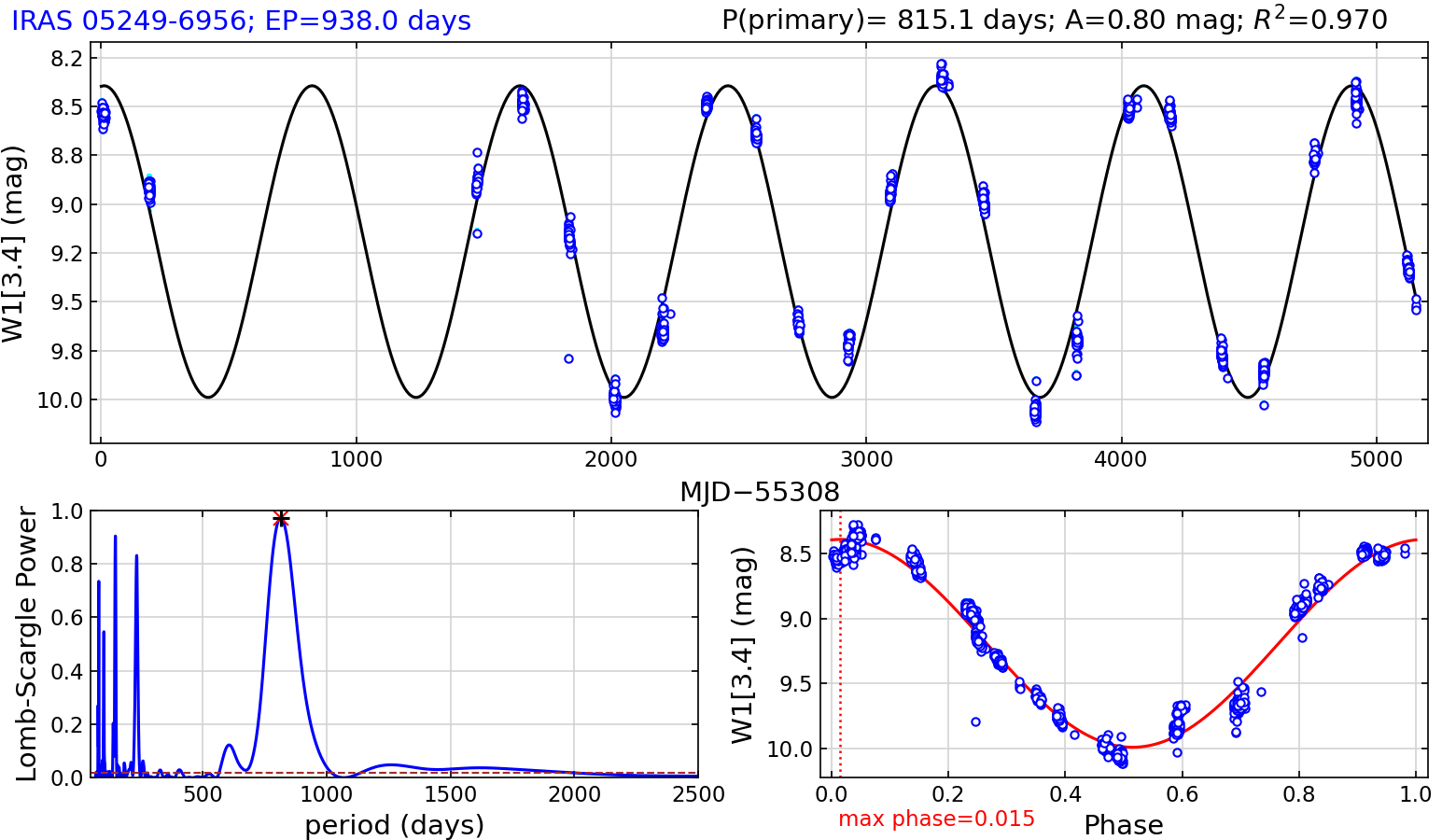}{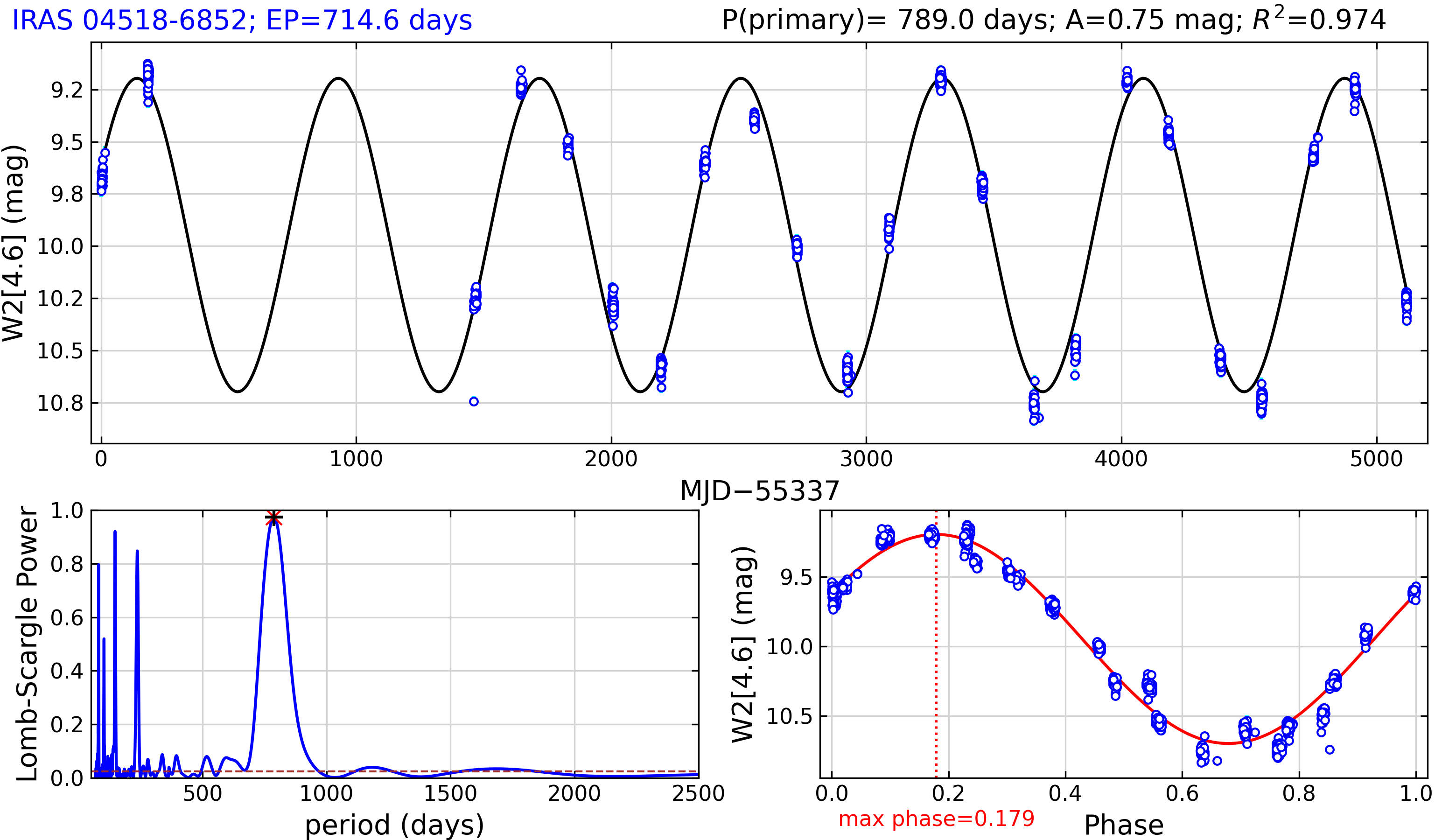}{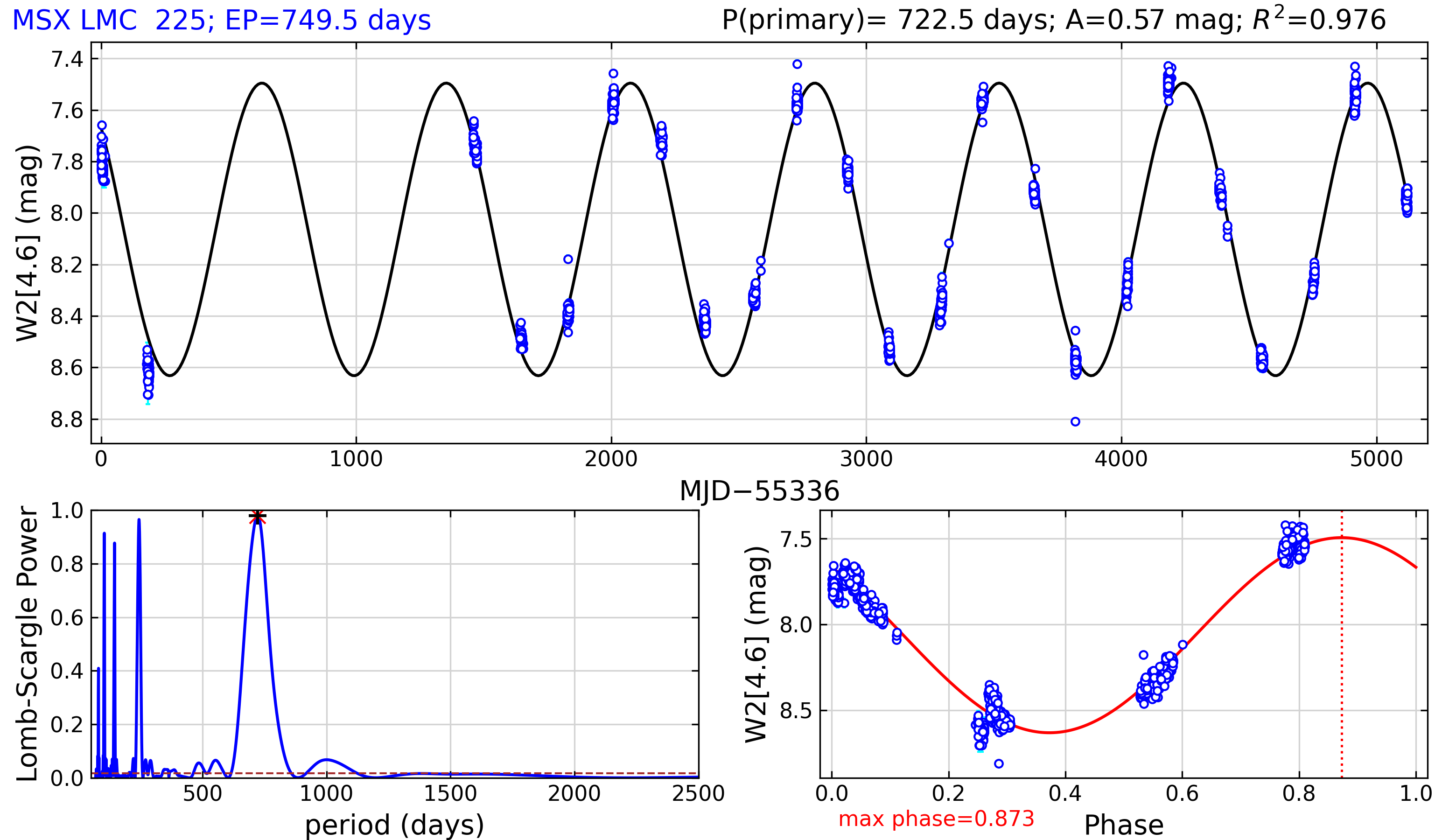}{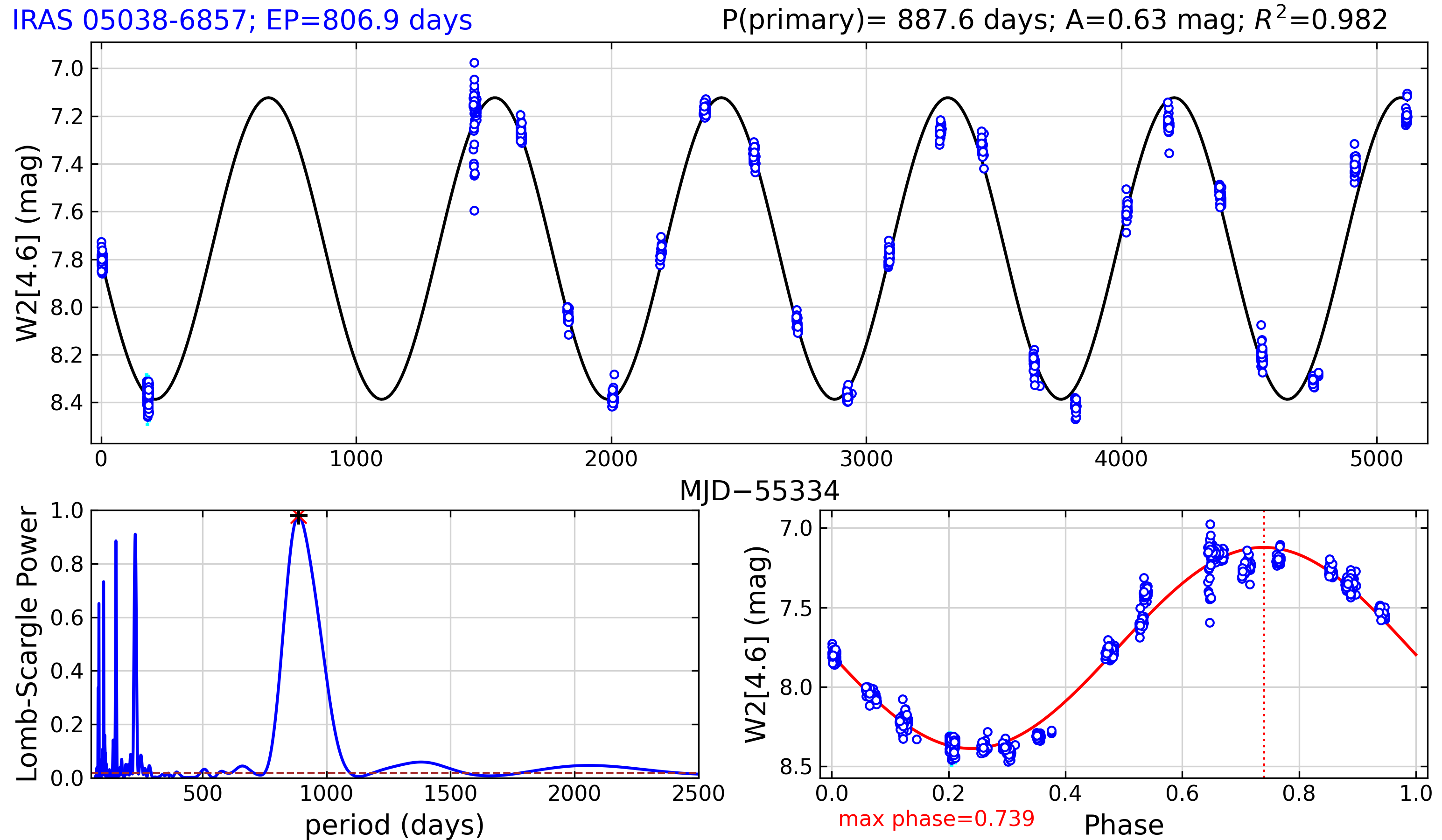}{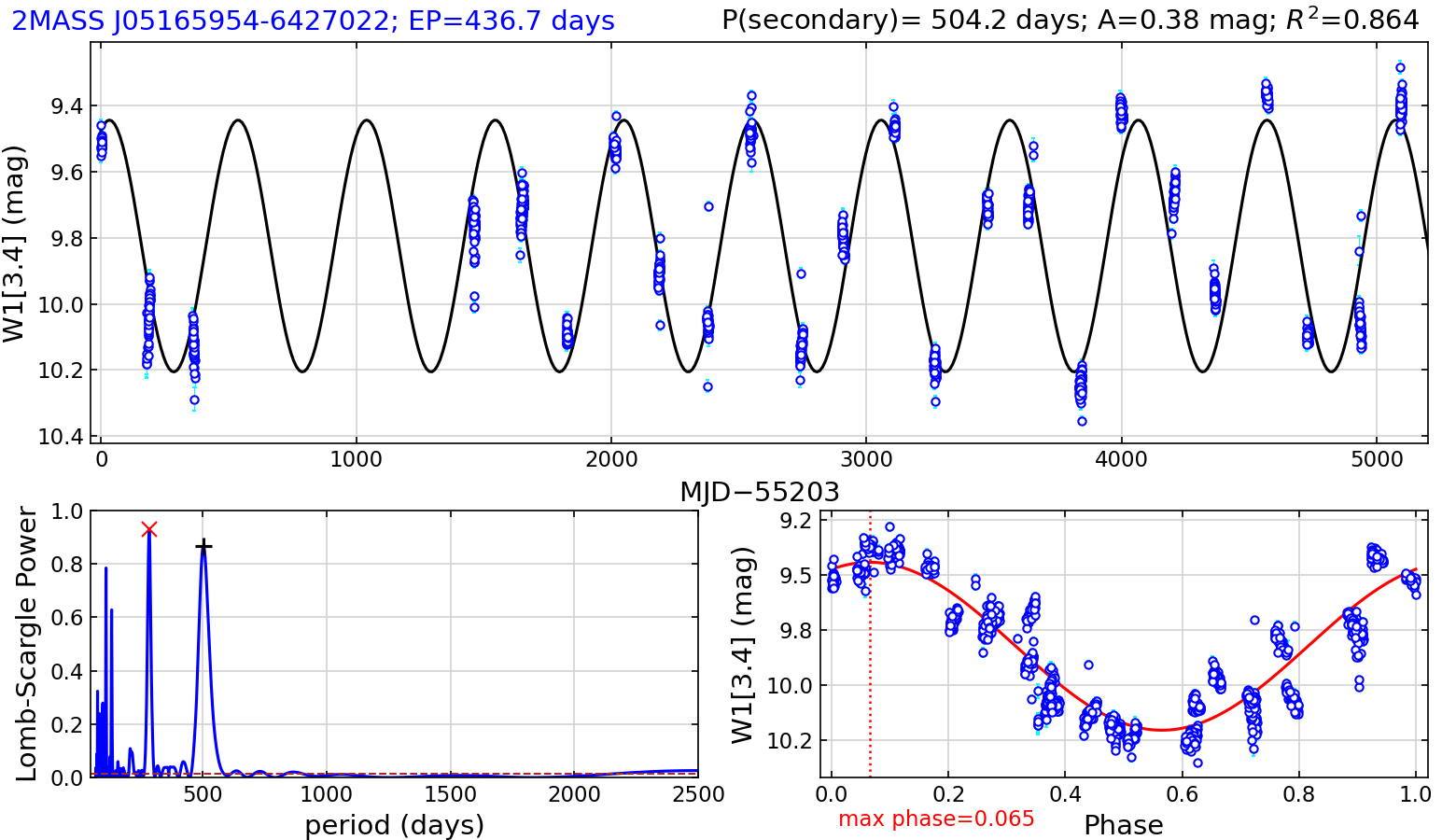}{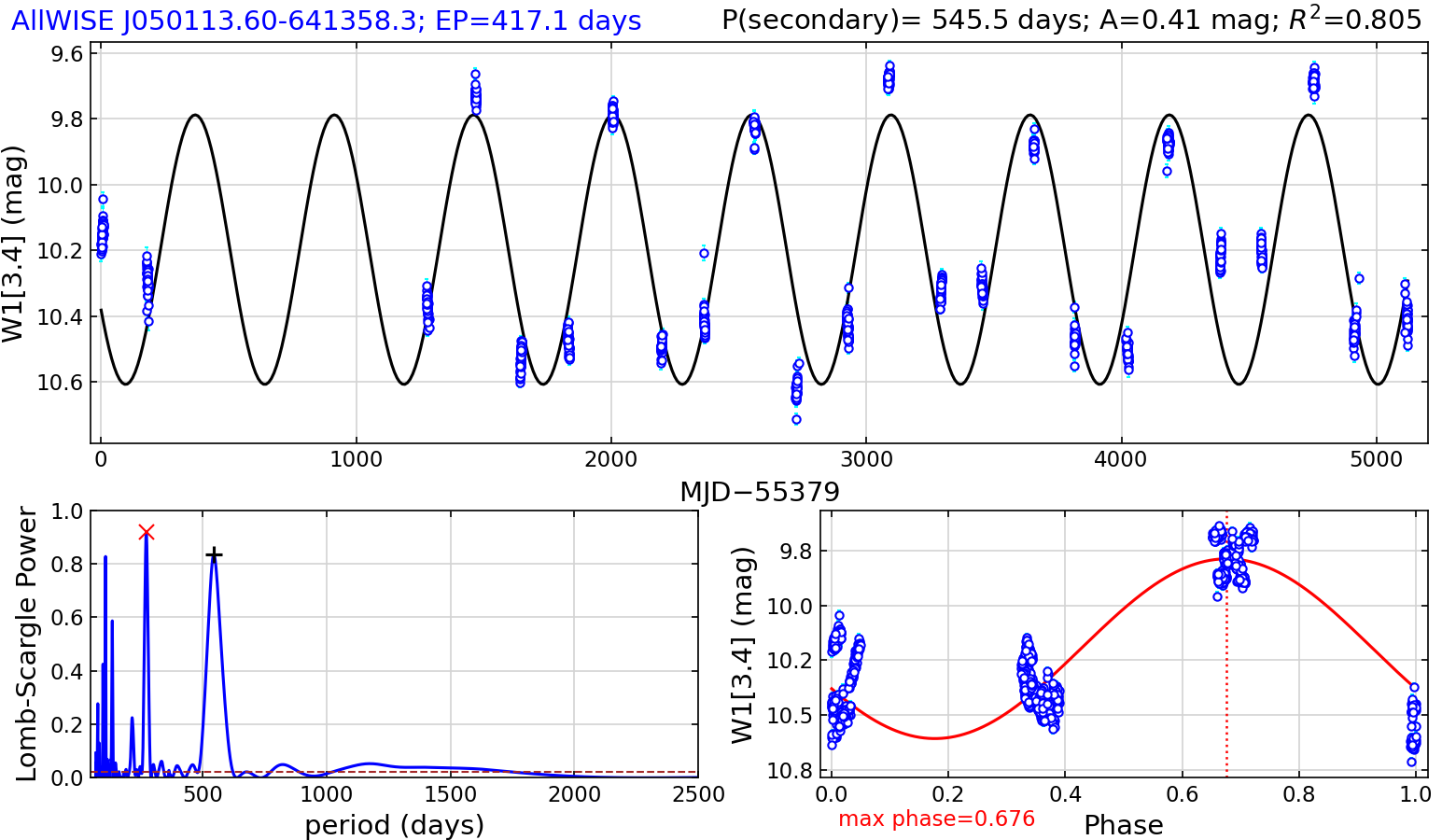}
\caption{Lomb-Scargle periodograms using WISE light curves for eight CAGB stars with unknown periods.
These objects are candidates for new Mira variables identified from WISE data.
The expected period (EP) denotes the period expected from their M(W3[12]) values using the PMR given in Equation~\eqref{eq:3} (see Section~\ref{sec:neo-mc}).
Note that the secondary peak is used for the last two objects.
See Section~\ref{sec:neo-a} for details.}
\label{f3}
\end{figure*}

\begin{figure*}
\centering
\smallplottwo{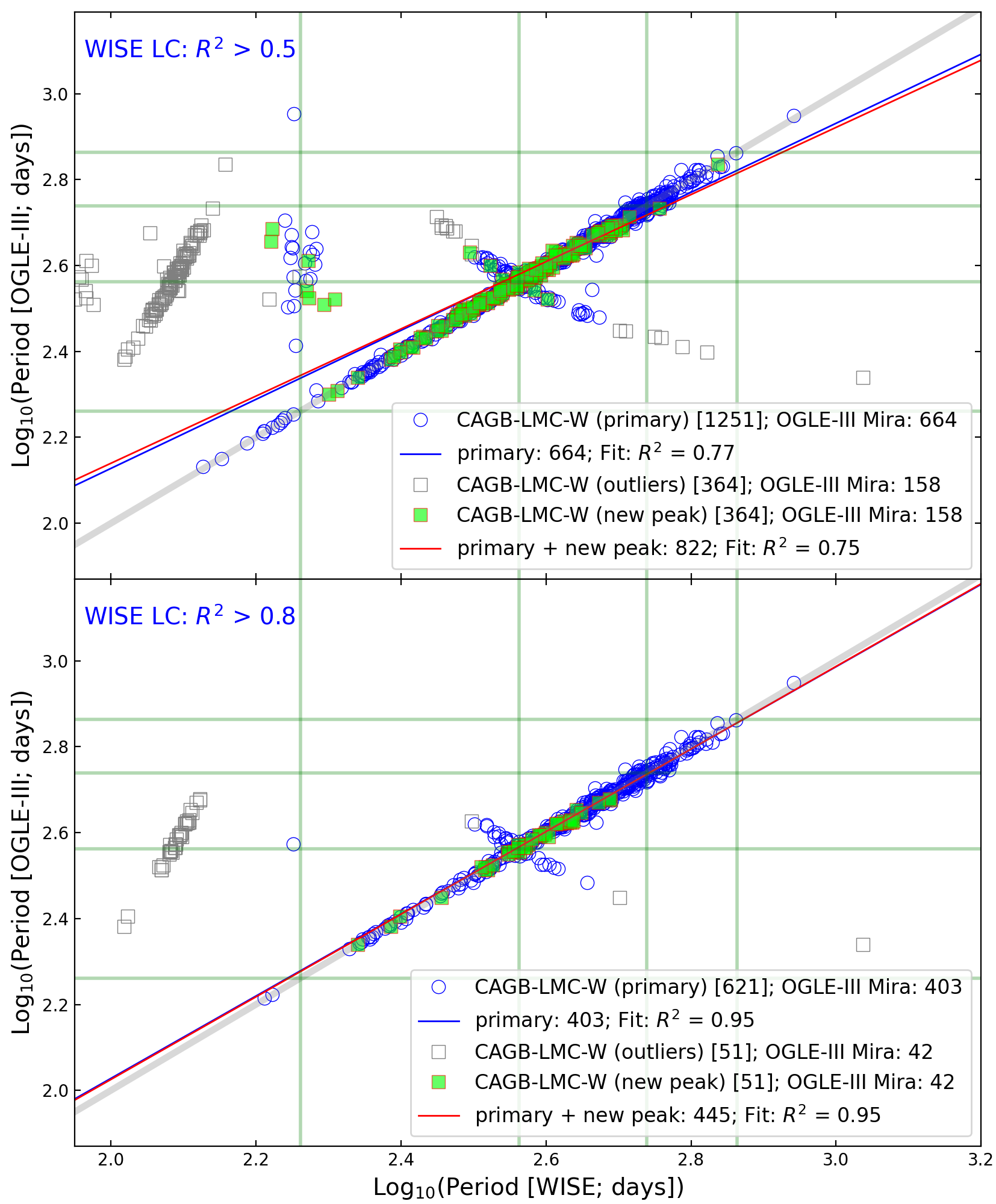}{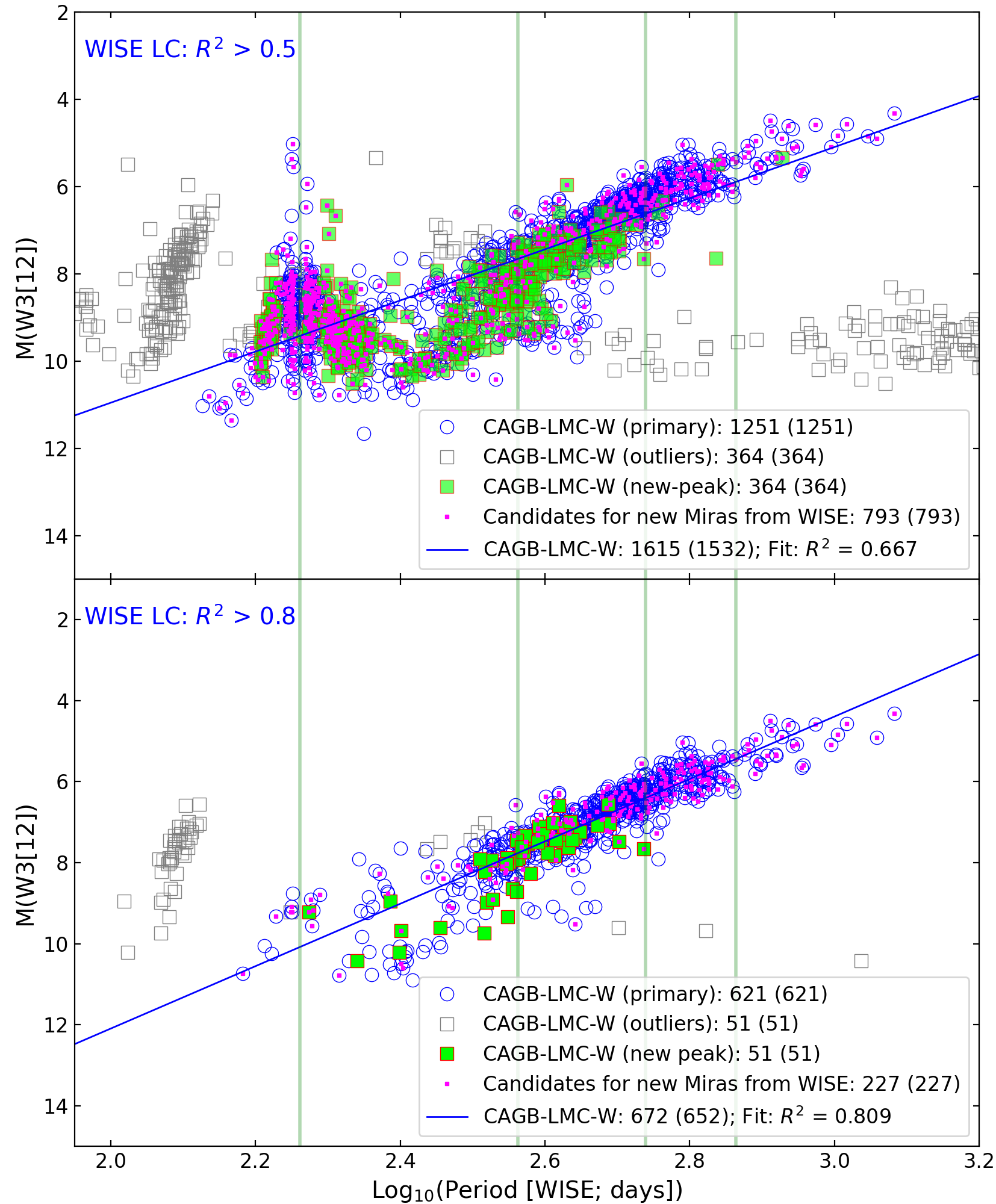}
\caption{Comparison of pulsation periods derived from OGLE-III and WISE data for known Miras from OGLE-III (left panels)
and period–magnitude relations for all Mira-like variables identified from WISE data
(right panels). The upper panels display objects with reliable WISE light curves ($R^2 > 0.5$),
while the lower panels show the objects with high-quality light curves ($R^2 > 0.8$).
See Section~\ref{sec:neo-mc} for details.}
\label{f4}
\end{figure*}

\subsection{Light curves from WISE data\label{sec:neo-a}}

For the 11,134 carbon stars in the LMC, we generated light curves using WISE data
and computed Lomb–Scargle periodograms. To identify reliable variability
parameters, we used $R^2$ rather than SNR, as $R^2$ is more effective for
distinguishing high-quality datasets, whereas SNR is better suited for screening
out low-quality data.

To ensure robust results, we applied stringent selection criteria, including a
minimum of 200 observed points, $R^2$ larger than 0.5, and a Lomb-Scargle power
exceeding 0.5. From the full sample of 11,134 carbon stars (including
CAGB-LMC-SAGE and CS-LMC-K objects), pulsation parameters were reliably derived
from WISE light curves for 1,615 stars (designated as CAGB-LMC-W objects; see
Table~\ref{tab:tab1}). Among these, 922 are identified as known Miras from
OGLE-III.

Among the 1,615 CAGB-LMC-W objects, 672 exhibit both an $R^2$ value greater than
0.8 and a Lomb-Scargle power exceeding 0.8, indicating pronounced Mira-like
variability. The number of such objects is given in parentheses in
Table~\ref{tab:tab1}. Primary periods were selected for 621 of them, while
secondary or tertiary peaks were chosen for the remaining 51 (see
Section~\ref{sec:neo-mc} for details). Of these 672 objects, 445 are previously
identified Miras from OGLE-III, and the remaining 227 are candidates for newly
identified Mira variables based on WISE data (see Table~\ref{tab:tab1}).

The top panel of Figure~\ref{f1} presents the Lomb-Scargle periodograms derived
from the OGLE-III light curve in the I[0.8] band, while the bottom panel shows
the WISE light curve in the W2[4.6] band (this work) for a CAGB star,
specifically the Mira variable MSX LMC 368 (OGLE-LMC-LPV-38456). Although the
I[0.8] and W2[4.6] band light curves exhibit similar variability patterns, the
W2[4.6] band displays a reduced amplitude. Unlike the OGLE-III periodogram, the
WISE periodogram shows multiple peaks of similar strength, likely due to the
fixed six-month observation schedule of WISE, as noted earlier.

Figure~\ref{f2} shows Lomb-Scargle periodograms for eight representative objects
from the 445 CAGB stars, which are known as Mira variables according to OGLE-III.
For the last two objects shown in the figure, the secondary period was selected.
The periods obtained from WISE data are very similar to those from OGLE-III.

Figure~\ref{f3} shows Lomb-Scargle periodograms for eight representative objects
from the 227 CAGB stars, which are not Miras according to OGLE-III, are
candidates for newly identified Mira variables based on WISE data. For the last
two objects shown in the figure, the secondary period was selected.

\subsection{Comparison of pulsation periods derived from OGLE-III and WISE data\label{sec:neo-mc}}

The left panels of Figure~\ref{f4} show the relation between OGLE-III periods and
the periods obtained from the WISE light curves for CAGB-LMC objects, that are
known as Miras from OGLE-III. The right panels of Figure~\ref{f4} show the PMRS
for the CAGB-LMC objects, for which pulsation parameters were reliably derived
from WISE light curves. The deviations tend to be more significant when the
OGLE-III or WISE periods align with integer multiples of WISE’s six-month
observation interval.

These deviations may result from the nature of the Lomb-Scargle periodogram,
which can produce multiple comparable peaks when applied to regularly sampled
data such as WISE observations taken every six months. Similar effects have also
been reported by \citet{vanderPlas2018} and \citet{suh2021}.

For the 1,615 objects with reliable WISE light curves (defined by $R^2 > 0.5$;
see the upper panels of Figure~\ref{f4}), the correlations show substantial
scatter. However, restricting the sample to the 672 objects with high-quality
light curves ($R^2 > 0.8$; see the lower panels of Figure~\ref{f4}) results in
significantly tighter correlations with reduced dispersion. For the remainder of
this study, we focus exclusively on these 672 high-quality WISE light curves.

Of the 672 high-quality objects, primary pulsation periods were determined for
621 sources: 605 from the CAGB-LMC-SAGE sample and 16 from the CS-LMC-K sample,
including 399 CAGB-LMC-SAGE Miras and 4 CS-LMC-K Miras (see
Table~\ref{tab:tab1}). These objects are shown as blue circle symbols.

For the remaining 51 objects, shown as empty black squares in Figure~\ref{f4}, 
the primary periods derived from WISE data deviate significantly from the 
expected period (EP). Of these, 42 are Miras previously identified by OGLE-III, 
for which the EP corresponds to the OGLE-III period. For the other 9 objects, 
which are not OGLE-III Miras, the EP is estimated from their M(W3[12]) value 
using the PMR given in Equation~\eqref{eq:3}, which is derived from OGLE-III 
Miras (see Section~\ref{sec:pmr}).

For all 51 outliers, secondary or tertiary periods were adopted instead and are 
shown as lime-filled squares in Figure~\ref{f4}, as these revised periods align 
more closely with the EPs from OGLE-III or the PMR. This adjustment significantly 
reduces the scatter.

Among the 672 high-quality objects ($R^2 > 0.8$), 227 objects (218 from
CAGB-LMC-SAGE and 9 from CS-LMC-K) are considered candidates for newly identified
Mira variables from WISE data and are represented by magenta dot symbols in
Figure~\ref{f4}.

\begin{figure}
\centering
\smallplot{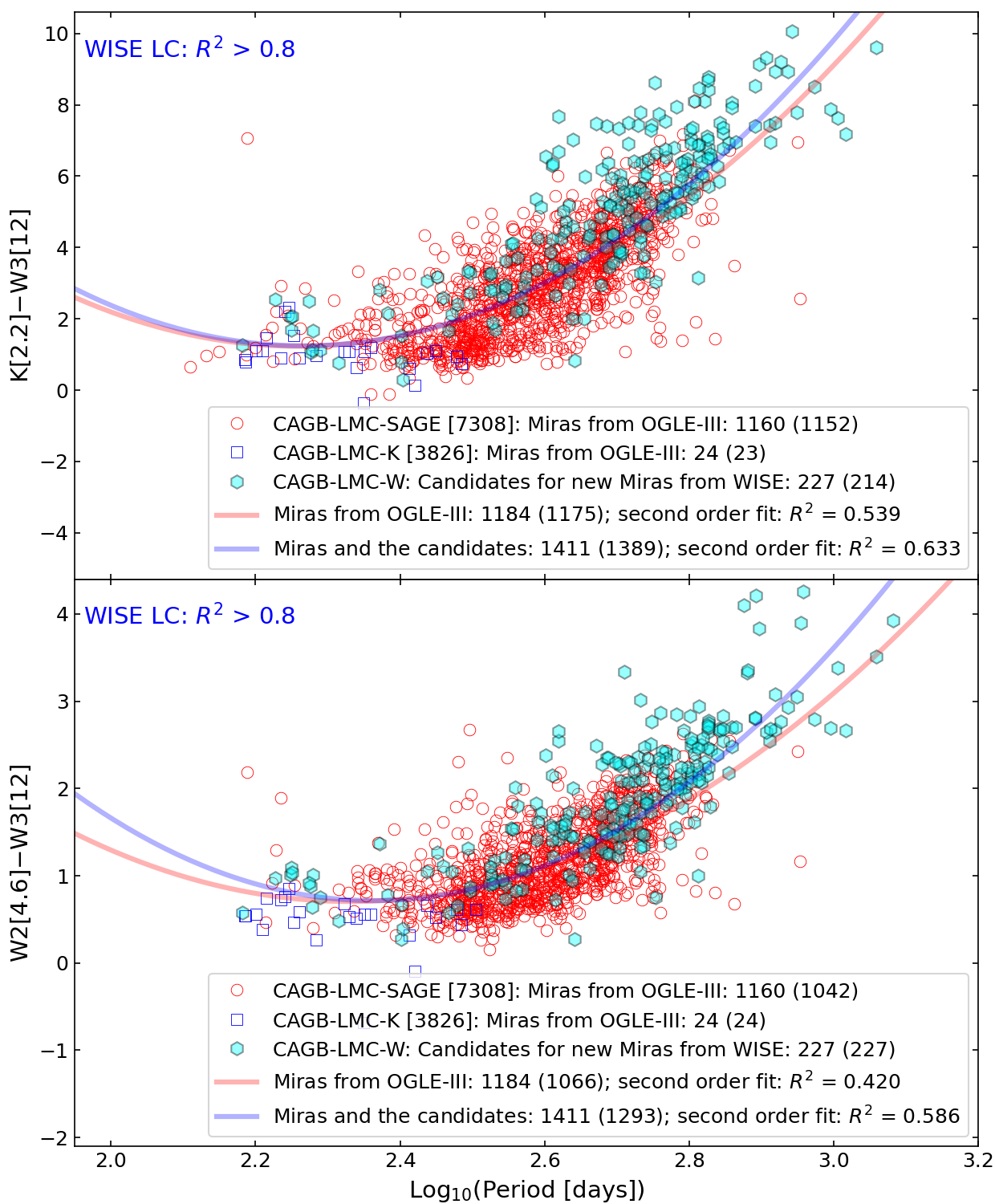}
\caption{Period–color relations for known Miras from OGLE-III and candidates for newly identified Miras
based on WISE light curves. Details are provided in Section~\ref{sec:pcr}.}
\label{f5}
\end{figure}

\section{Properties of Mira variables in the LMC\label{sec:pul}}

AGB stars typically exhibit long-period, large-amplitude pulsations. It is widely
accepted that as AGB stars evolve (or for those with higher initial masses), they
tend to show increased pulsation amplitudes, longer periods, and enhanced
mass-loss rates (e.g., \citealt{debeck2010}; \citealt{suh2021}).

\subsection{Period-color relations\label{sec:pcr}}

Figure~\ref{f5} presents the period–color relations, specifically plotting the
K[2.2]$-$W3[12] and W2[4.6]$-$W3[12] colors against pulsation periods for
carbon-rich Mira variables in the LMC. Despite considerable scatter, a clear
trend emerges, indicating a correlation between the IR color and pulsation period
for these stars.

As AGB stars evolve, those with longer pulsation periods typically develop
thicker circumstellar dust envelopes, leading to redder IR colors. Compared to
CAGB stars in the Milky Way (see \citealt{suh2020}), LMC Miras demonstrate a
stronger period–color correlation. This difference may stem from selection
effects, as many dust-enshrouded Galactic Miras are absent from the CAGB catalog
(e.g., \citealt{suh2020}). Galactic Mira samples are often based on optical data,
which suffer from heavy extinction in the Galactic disk, potentially biasing the
sample against the most dust-obscured stars.

\begin{figure*}
\centering
\smallplotfour{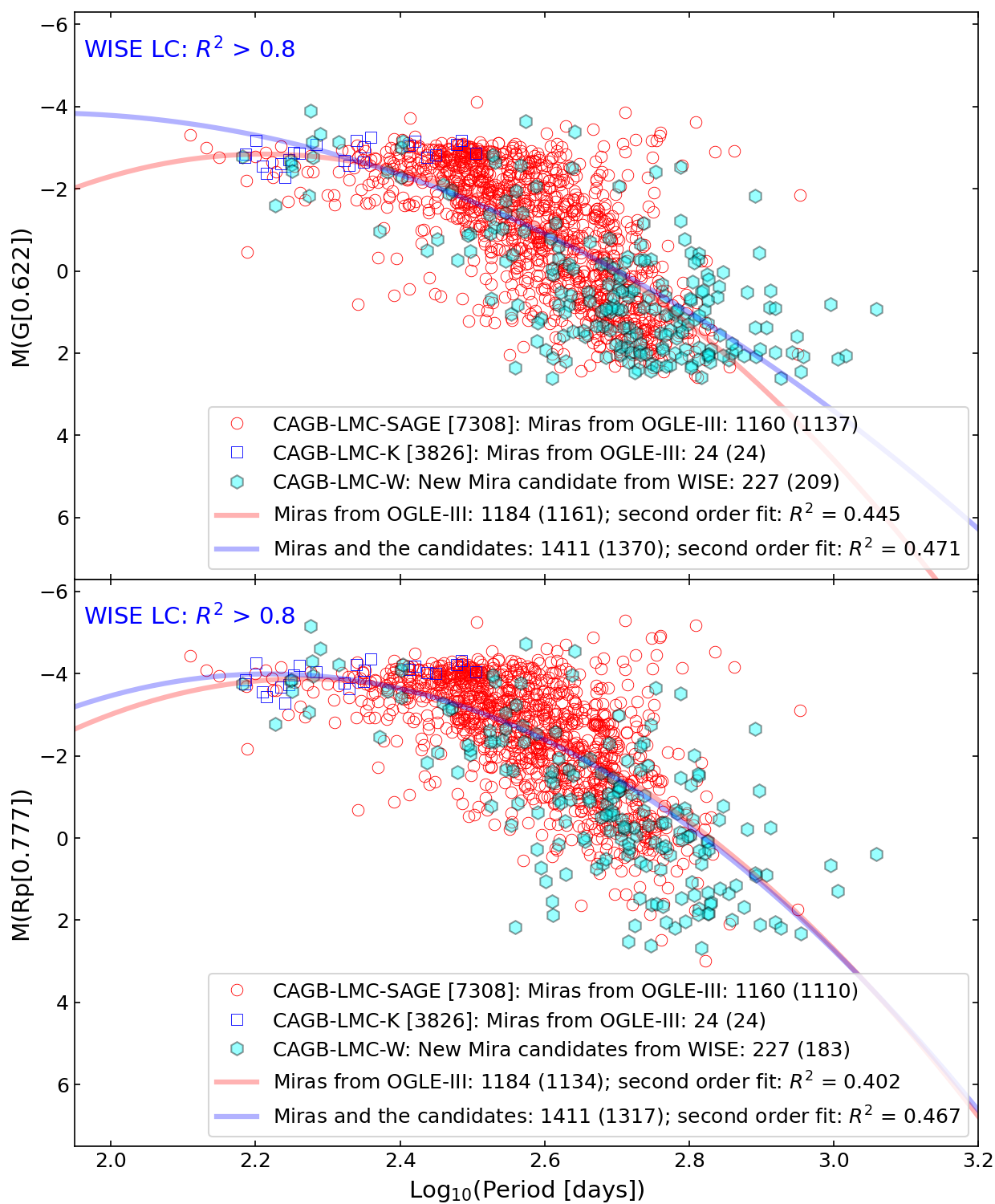}{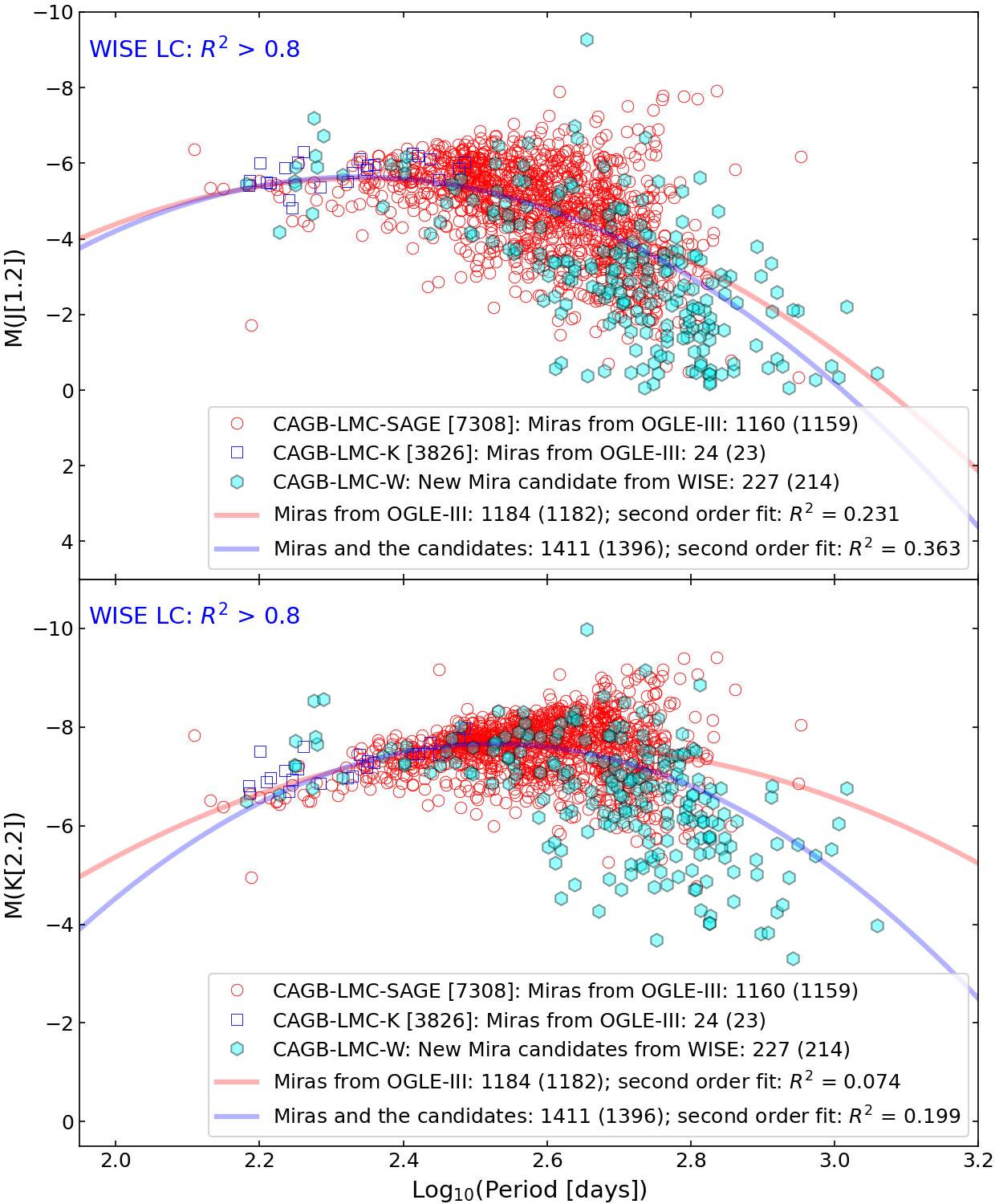}{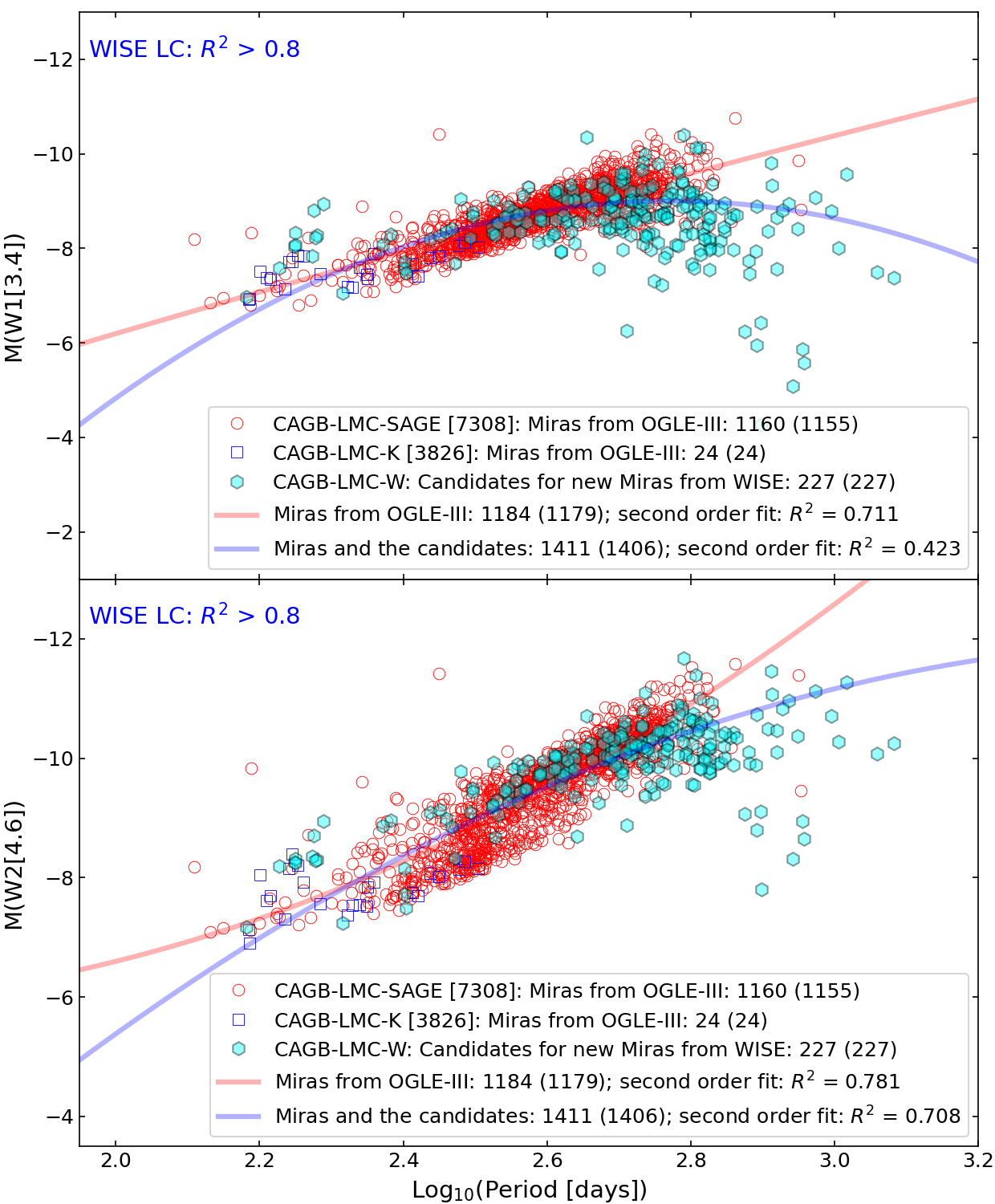}{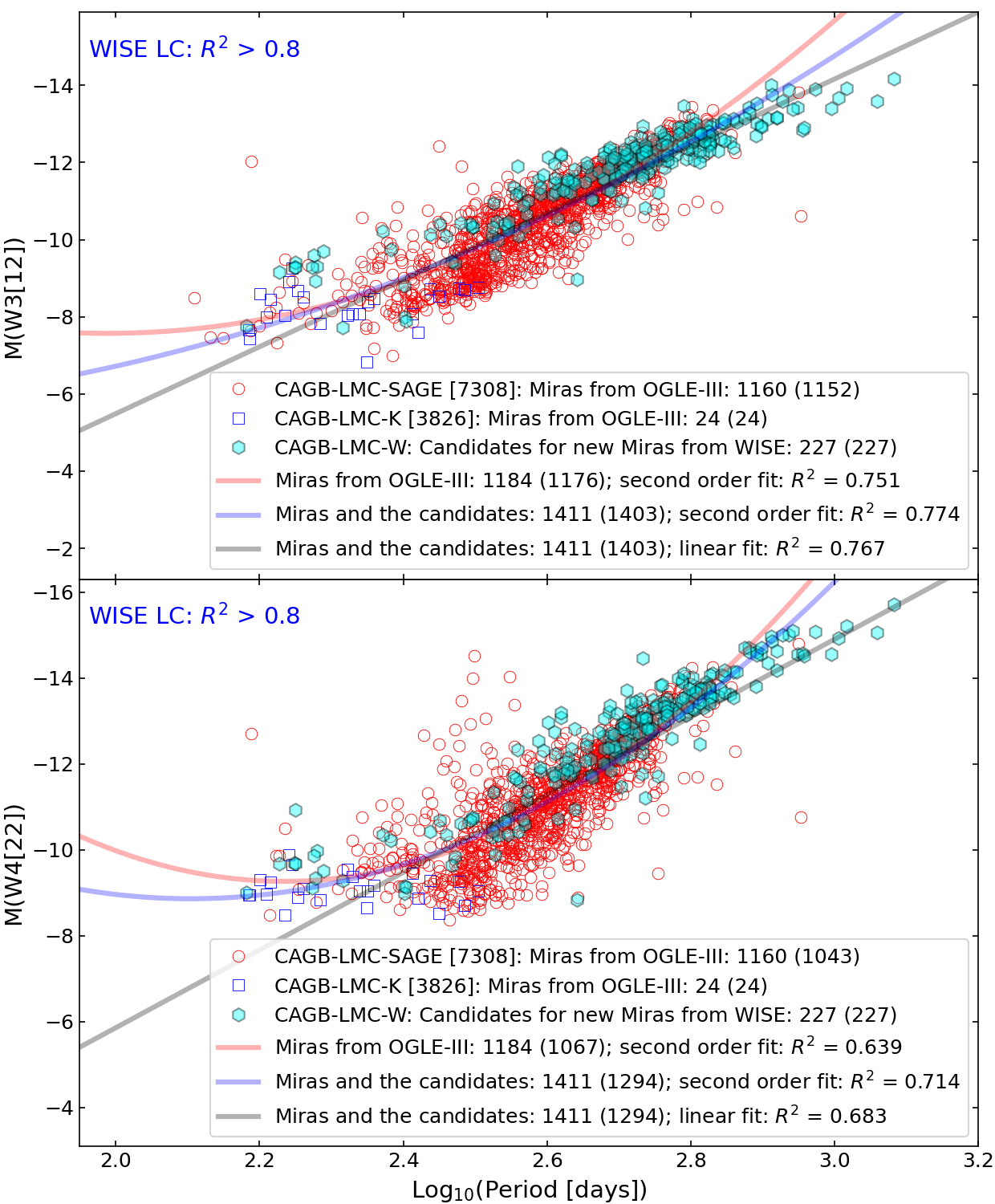}
\caption{Period-magnitude relations for known Miras from OGLE-III and candidates for newly identified Miras
based on WISE light curves.
The total number of objects for each subgroup is indicated, with the number in parentheses
representing the count of plotted objects with available observational data.
See Section~\ref{sec:pmr} for details.}
\label{f6}
\end{figure*}

\begin{figure*}
\centering
\smallploteight{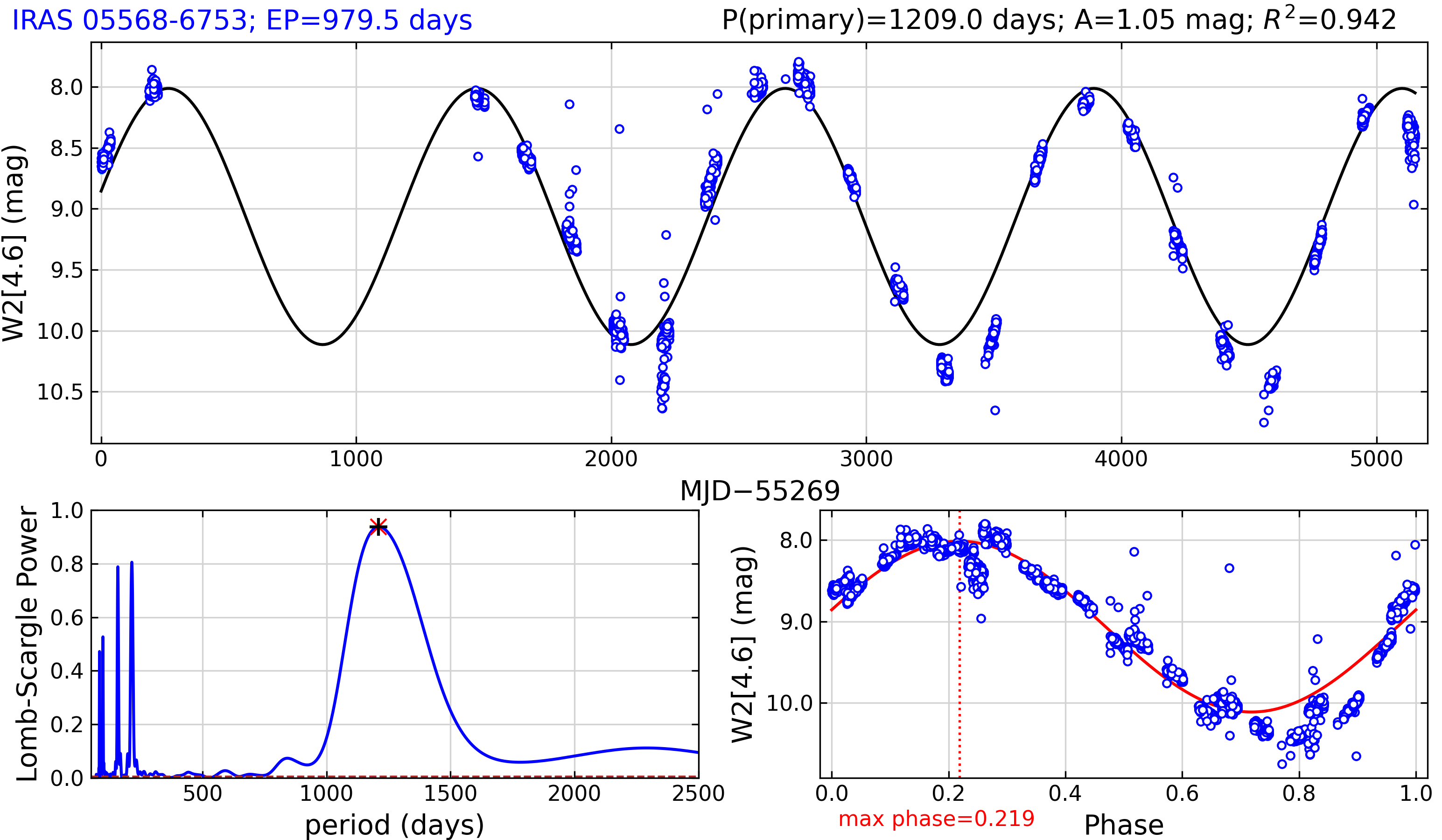}{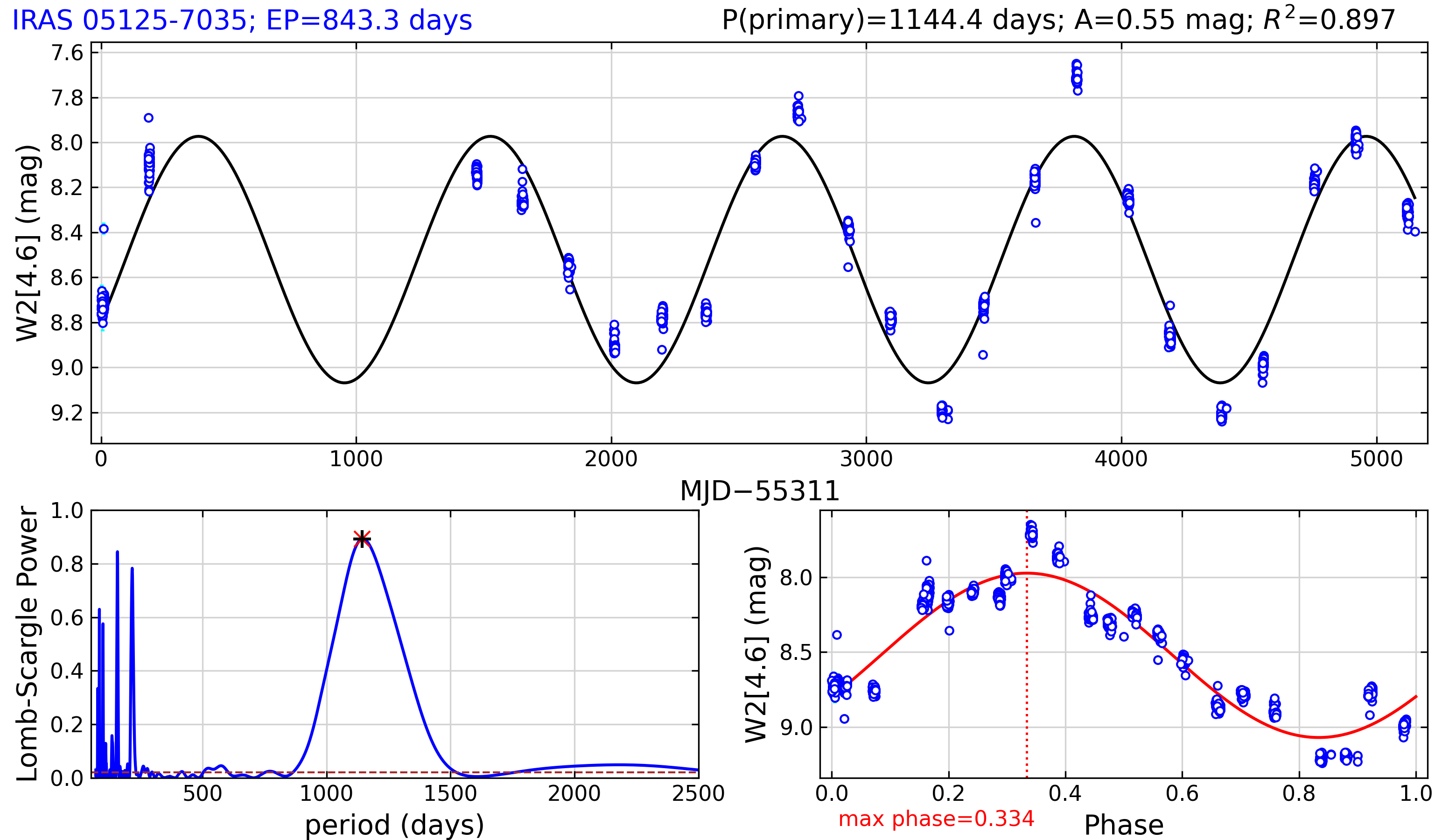}{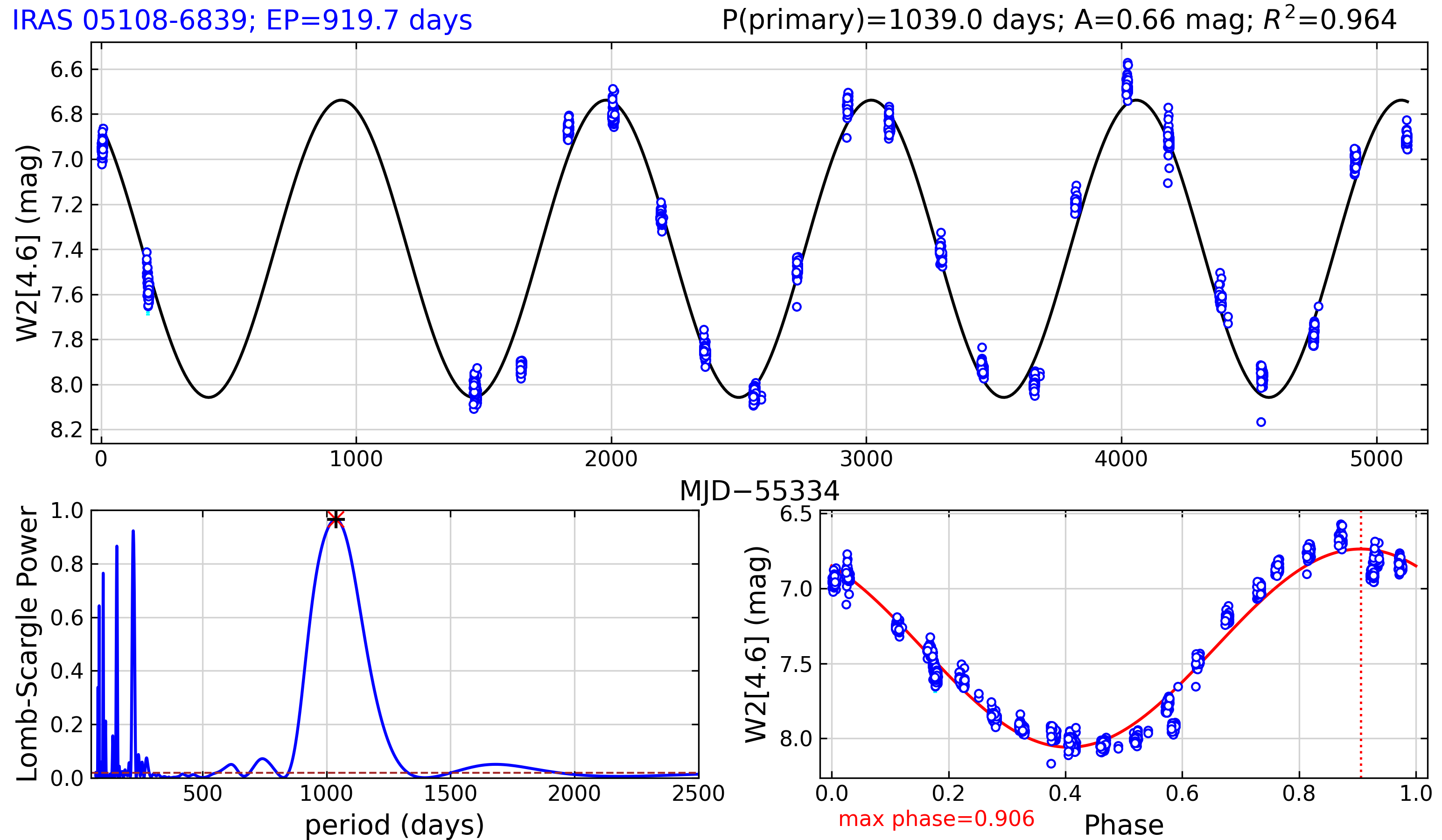}{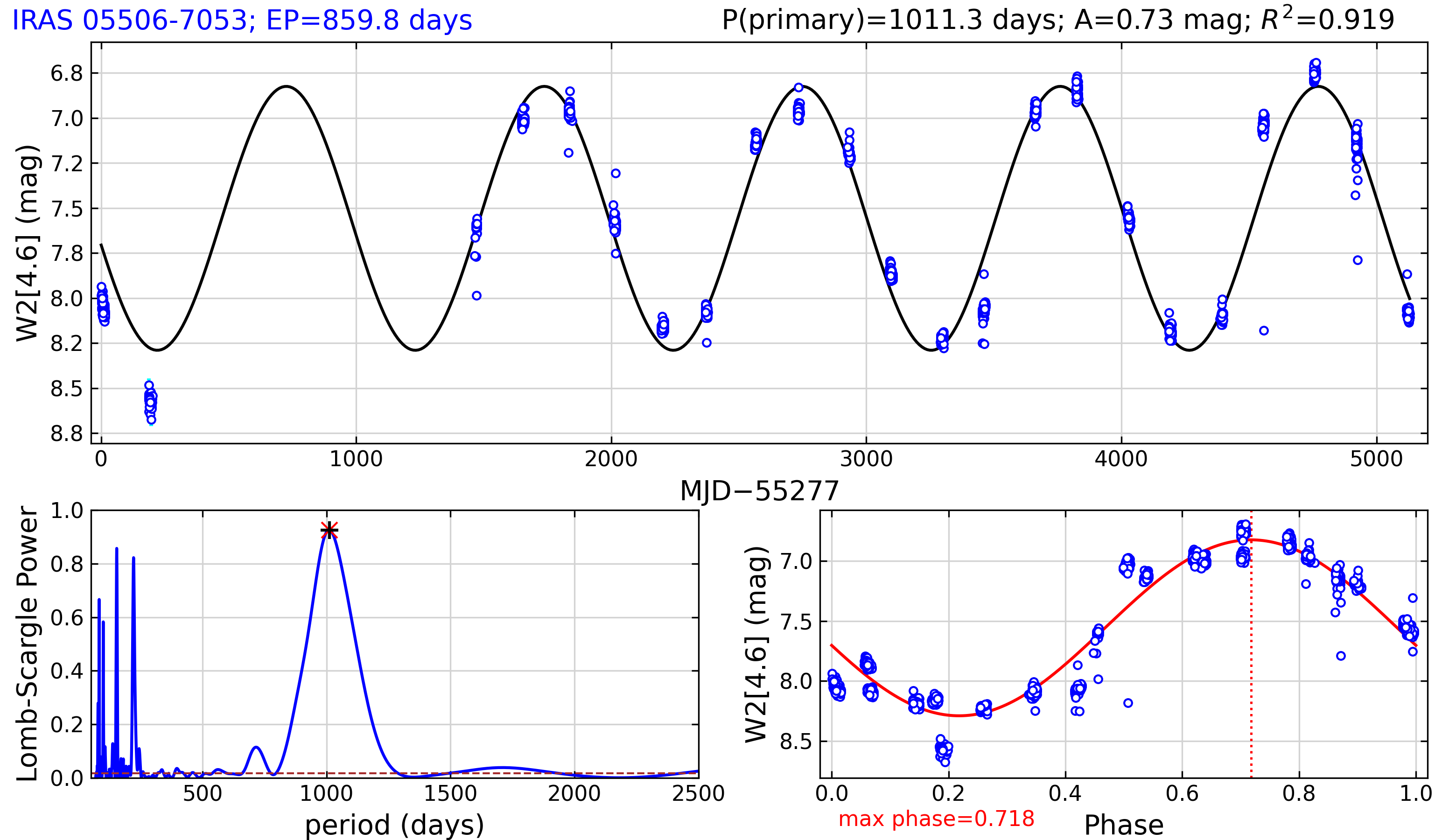}{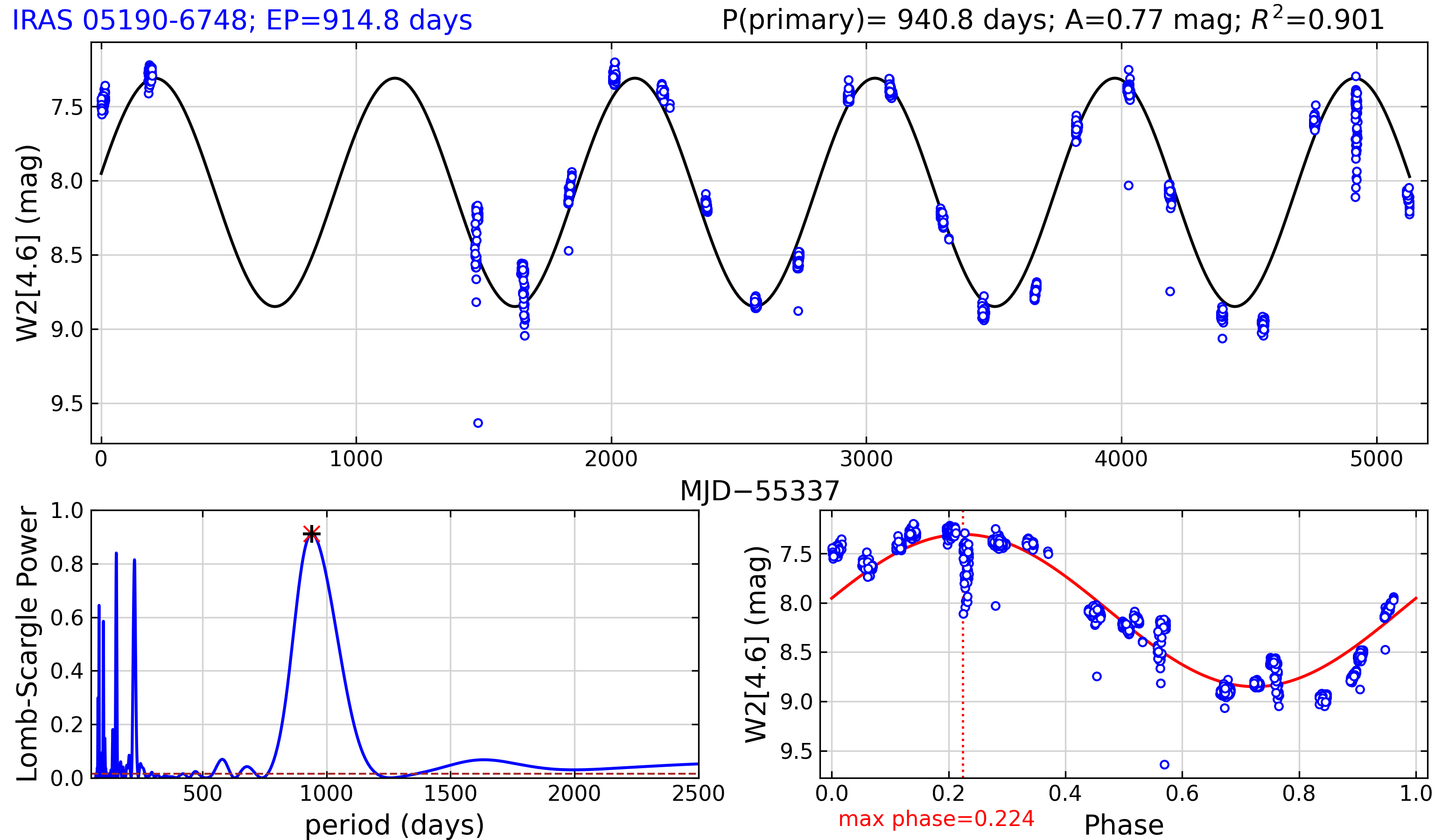}{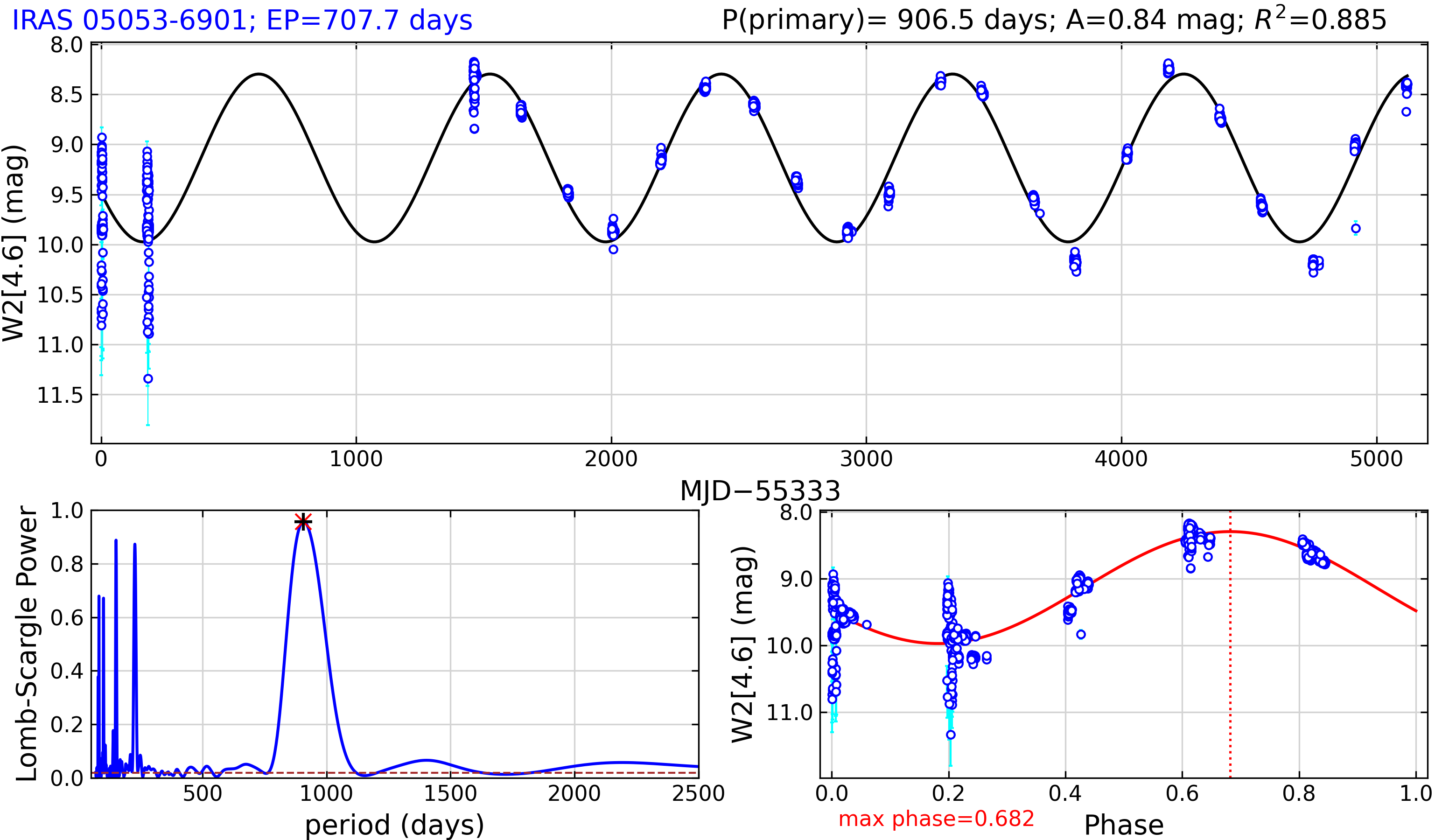}{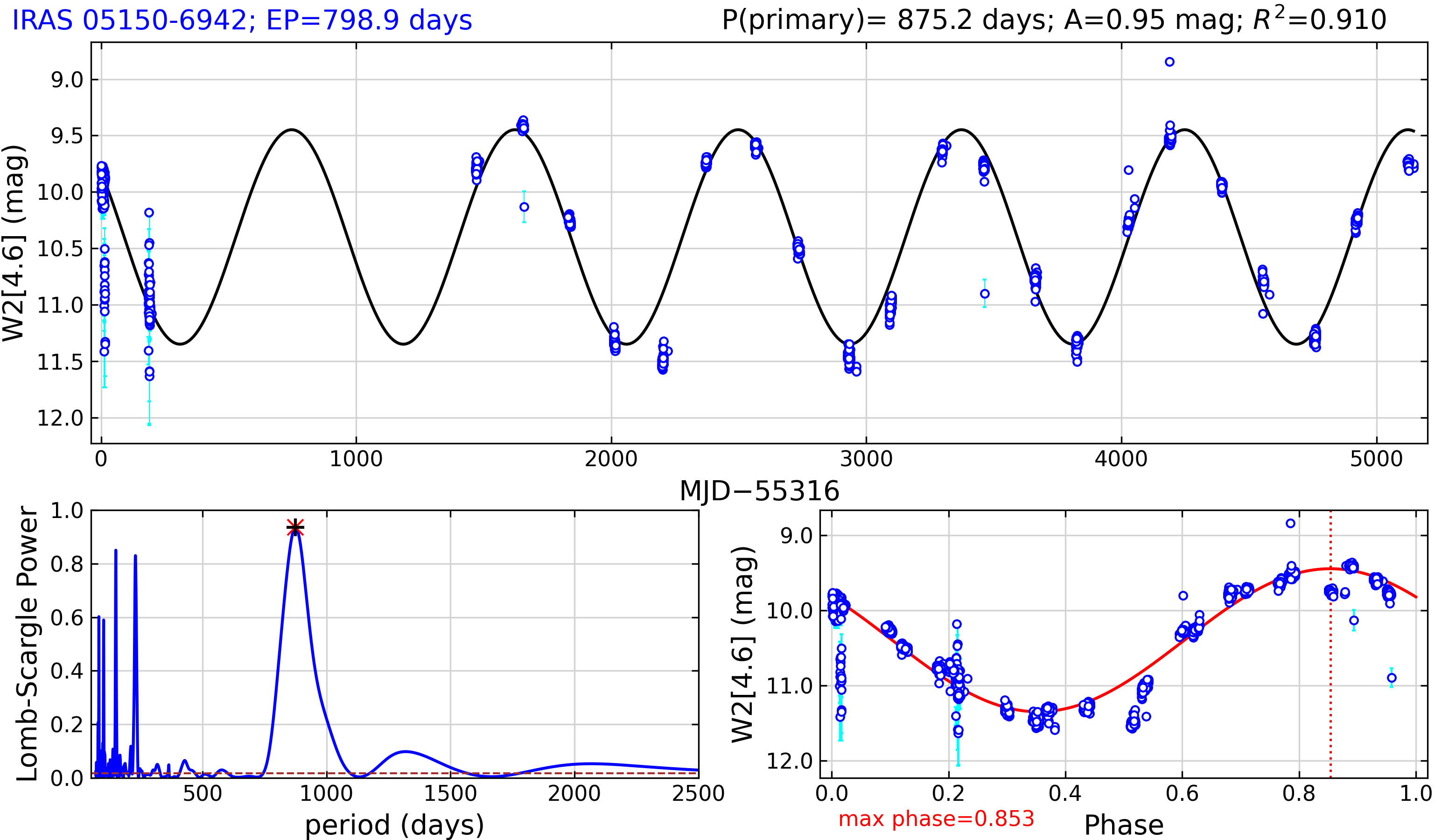}{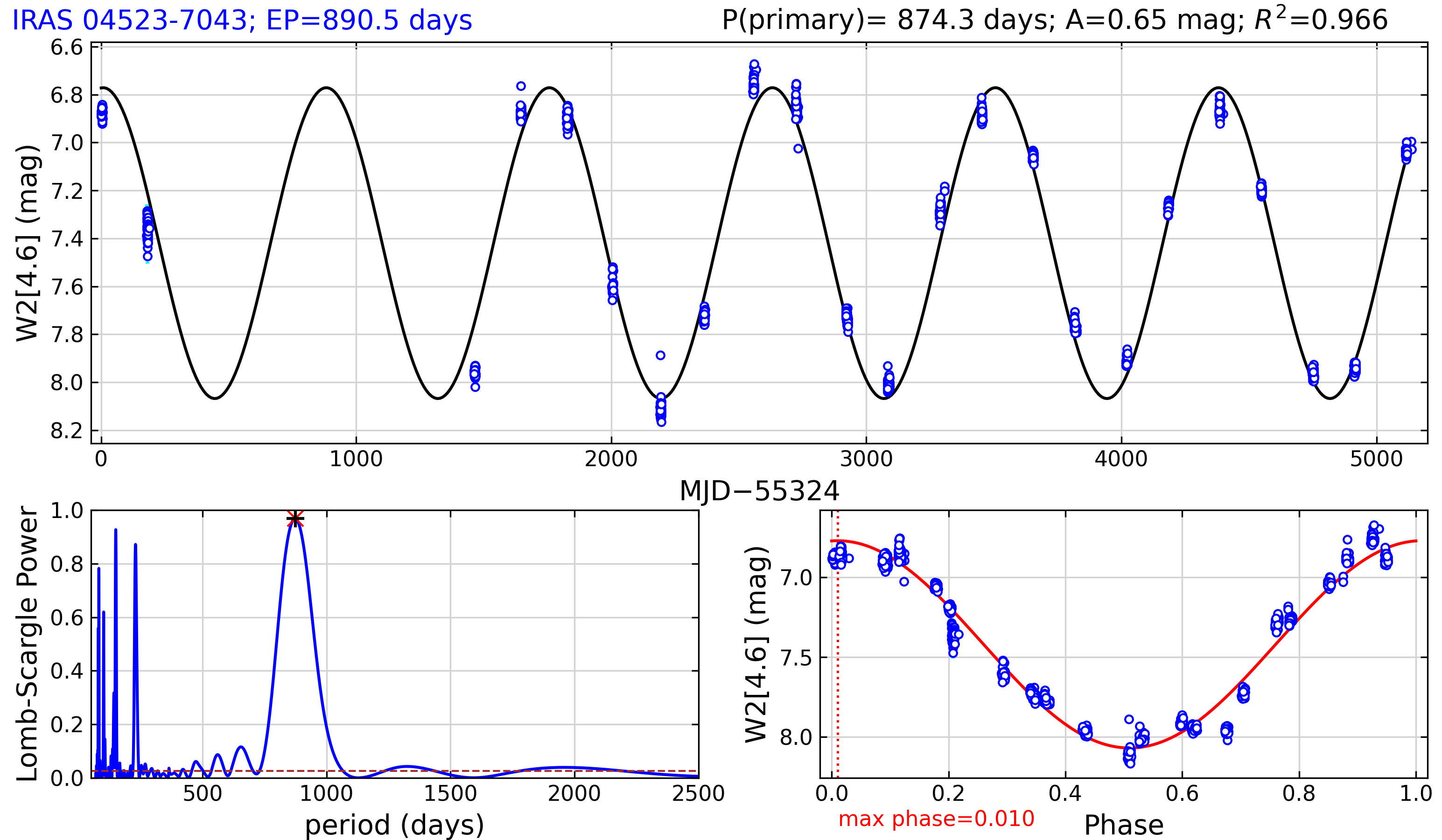}
\caption{Lomb–Scargle periodograms based on WISE light curves for eight CAGB stars with periods longer than 800 days.
These objects are candidates for new Mira variables identified from WISE data.
The expected period (EP) denotes the period expected from their M(W3[12]) values using the PMR given in Equation~\eqref{eq:3} (see Section~\ref{sec:neo-mc}).
Note that the primary period is used for all eight of these objects.
Details are provided in Section~\ref{sec:pmr}.}
\label{f7}
\end{figure*}

\subsection{Period-magnitude relations\label{sec:pmr}}

Studying the period–magnitude relation (PMR) of AGB stars is more straightforward 
in the Magellanic Clouds than in our Galaxy, as their distances are relatively 
uniform. Miras in the LMC consistently exhibit a well-defined single-sequence PMR 
that is clearer in the MIR bands than at shorter wavelengths (e.g., 
\citealt{sus09}; \citealt{suh2020}; \citealt{iwanek2021}).

Figure~\ref{f6} shows the PMRs for Mira variables in the LMC, including those 
identified by OGLE-III and newly identified candidates from WISE data, across 
visual and IR bands. To calculate the absolute magnitudes in the visual and IR 
bands, we adopt a distance of 50 kpc to the LMC, a commonly used approximation 
based on the precise measurement of 49.59 $\pm$ 0.54 kpc by 
\citet{pietrzynski2019}.

Although the relations exhibit substantial scatter in the visual (G[0.622] and
Rp[0.777]) and 2MASS (J[1.2] and K[2.2]) bands, the Miras display a strong
second-order correlation at longer wavelengths (i.e., in the WISE bands). We find
that Mira variables in the LMC show relatively high coefficients of determination
($R^2$ $>$ 0.7) for second-order or linear fits across the 3–24 $\mu$m range. In
the W3[12] and W4[22] bands, the PMRs are nearly linear, as the linear fits are
comparable in strength to the second-order fits.

The OGLE-III Mira sample, based on the I[0.8] band, is limited into CAGB stars 
with thin dust envelopes and relatively short pulsation periods. In contrast, the 
new Mira candidates identified from WISE data in the W1[3.4] and W2[4.6] bands 
may include CAGB stars with thicker dust envelopes and longer poulsation periods. 
Considering both the OGLE-III Miras and the WISE Mira candidates, the PMR in the 
W3[12] band shows the strongest correlation among those presented in 
Figure~\ref{f6}. Therefore, it can be regarded as the most reliable PMR.

Considering both the OGLE-III Miras and the WISE Mira candidates, the PMR in the
W3[12] band (see the lower-right panel of Figure~\ref{f6}) shows similarly good
fits for both a second-order and a linear model. The PMR derived from the
second-order fit is:
\begin{equation}
\mathrm{M(W3[12])} = -3.86 (\log_{10} P)^{2} + 11.3 \log_{10} P - 13.8 \pm 0.65,\label{eq:1}
\end{equation}
where $P$ is the pulsation period in days. For the linear fit, the PMR is:
\begin{equation}
\mathrm{M(W3[12])} = -8.68 \log_{10} P + 11.9 \pm 0.66.\label{eq:2}
\end{equation}
We expect that these PMRs may serve as valuable references for studying carbon 
stars in other galaxies, including the Milky Way. When considering only the Miras 
identified by OGLE-III, the linear-fit PMR in the W3[12] band becomes: 
\begin{equation}
\mathrm{M(W3[12])} = -8.88 \log_{10} P + 12.4 \pm 0.67.\label{eq:3}
\end{equation}

The shape of the PMRs shown in Figure~\ref{f6} (and the PCRs in Figure~\ref{f5}) 
is primarily influenced by the long-period objects ($\log_{10} P 
> 2.9$) newly discovered in this study from WISE data. These long-period objects,
among the 227 candidates for new Mira variables identified from WISE data, are
characterized by redder IR colors and thicker dust envelopes. Figure~\ref{f7}
presents Lomb–Scargle periodograms for eight representative long-period objects,
arranged in descending order of period.

\section{Theoretical Dust Shell Models\label{sec:models}}

On all of the CMDs and 2CDs in Figures~\ref{f7}, theoretical model tracks for
CAGB stars are plotted to be compared with the observations. To calculate
theoretical model SEDs for CAGB stars, we utilize radiative transfer models
designed for spherically symmetric dust shells surrounding central stars. We
employ the radiative transfer code
RADMC-3D\footnote{\url{http://www.ita.uni-heidelberg.de/~dullemond/software/radmc-3d/}},
applying the same methodologies as employed by \citet{suh2024}.

For CAGB stars, we adopt the same modeling approach as described in
\citet{suh2024}. All models assume a dust density profile following a continuous
power law ($\rho \propto r^{-2}$). The dust condensation temperature ($T_c$) is
fixed at 1000 K, which determines the inner boundary of the dust shell. The outer
boundary is set at $10^4$ times the inner radius. Dust grains are assumed to be
spherical with a uniform radius of 0.1 $\mu$m. We define the dust optical depth
($\tau_{10}$) at a reference wavelength of 10 $\mu$m.

For the dust composition, we use the optical constants of amorphous carbon (AMC)
grains from \citet{suh2000}. We compute seven models with optical depths
$\tau_{10}$ = 0.0001, 0.01, 0.1, 0.5, 1, 2, and 4. The stellar blackbody
temperature is set to 3000 K for $\tau_{10}$ = 0.0001, 2700 K for 0.01, 2500 K
for 0.1 and 0.5, 2200 K for 1, and 2000 K for $\tau_{10}$ = 2 and 4.

\begin{figure*}
\centering
\smallploteight{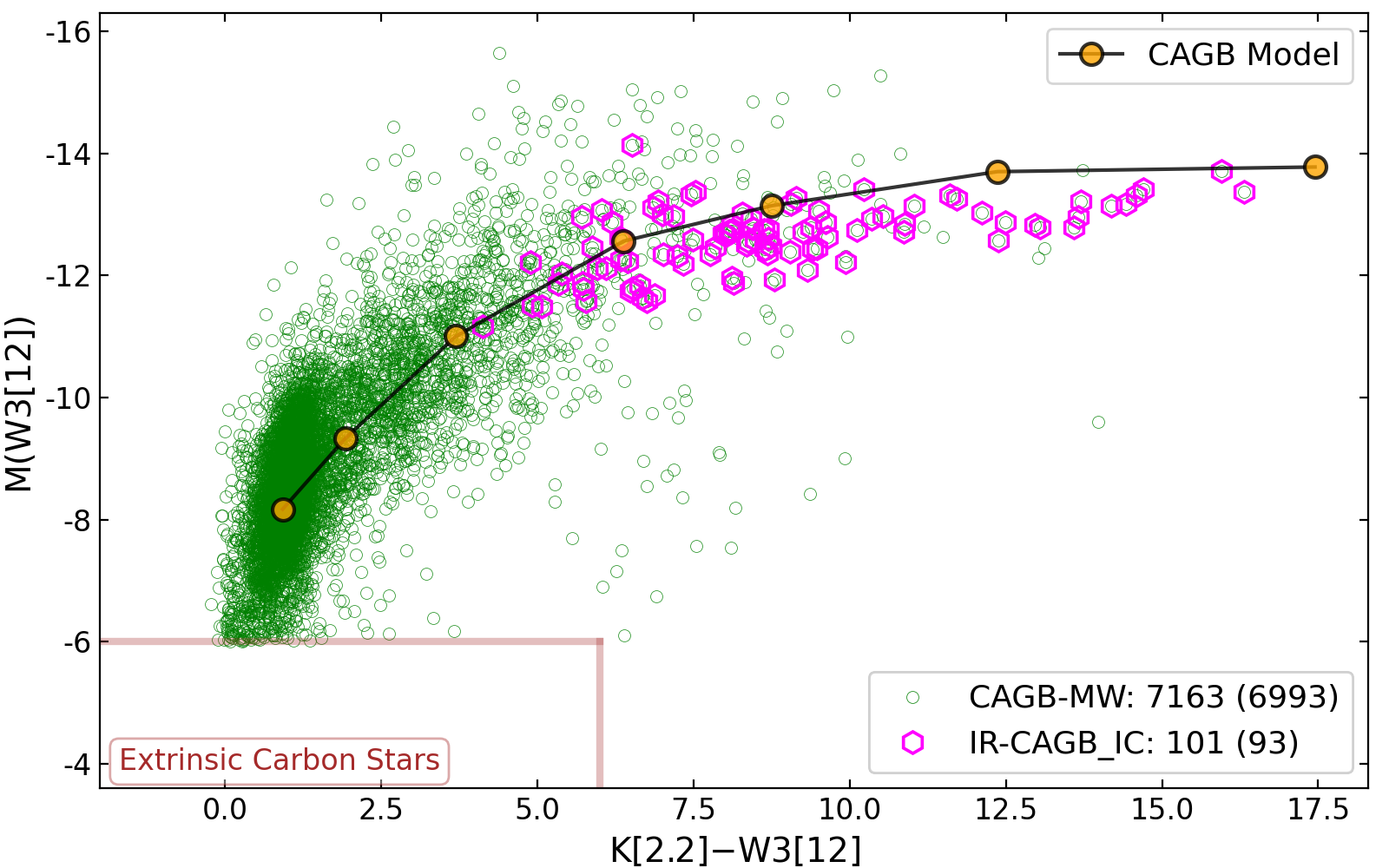}{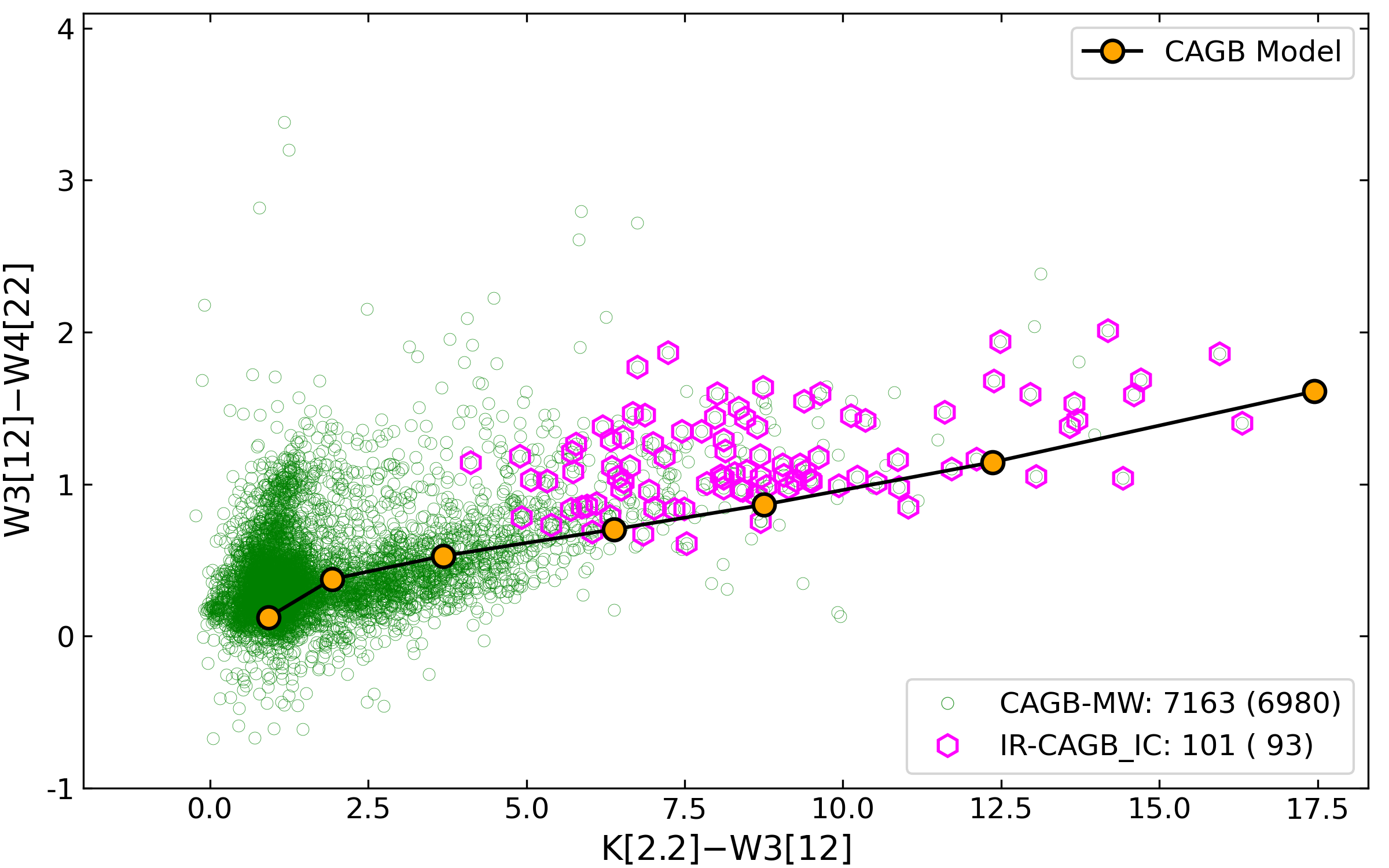}{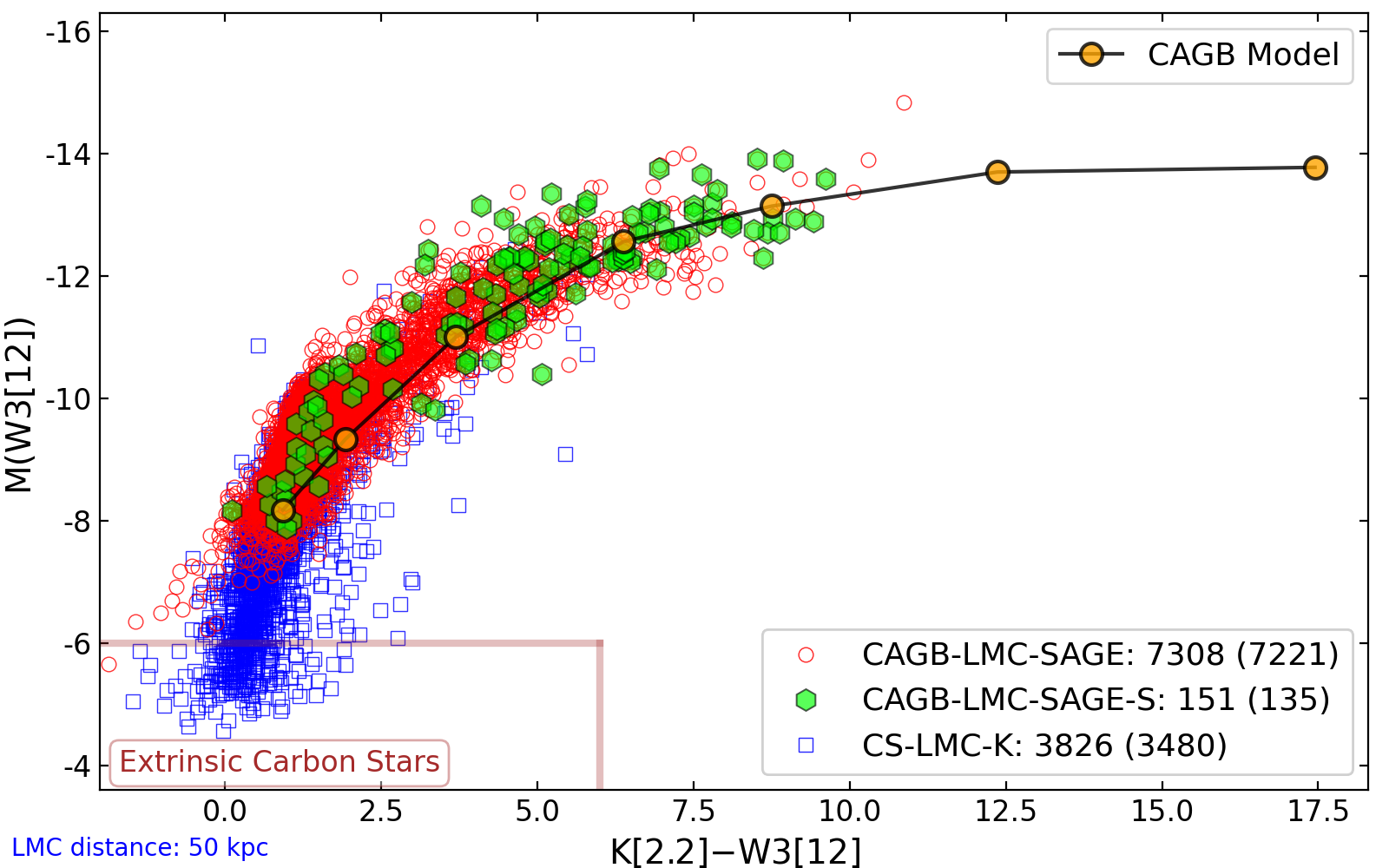}{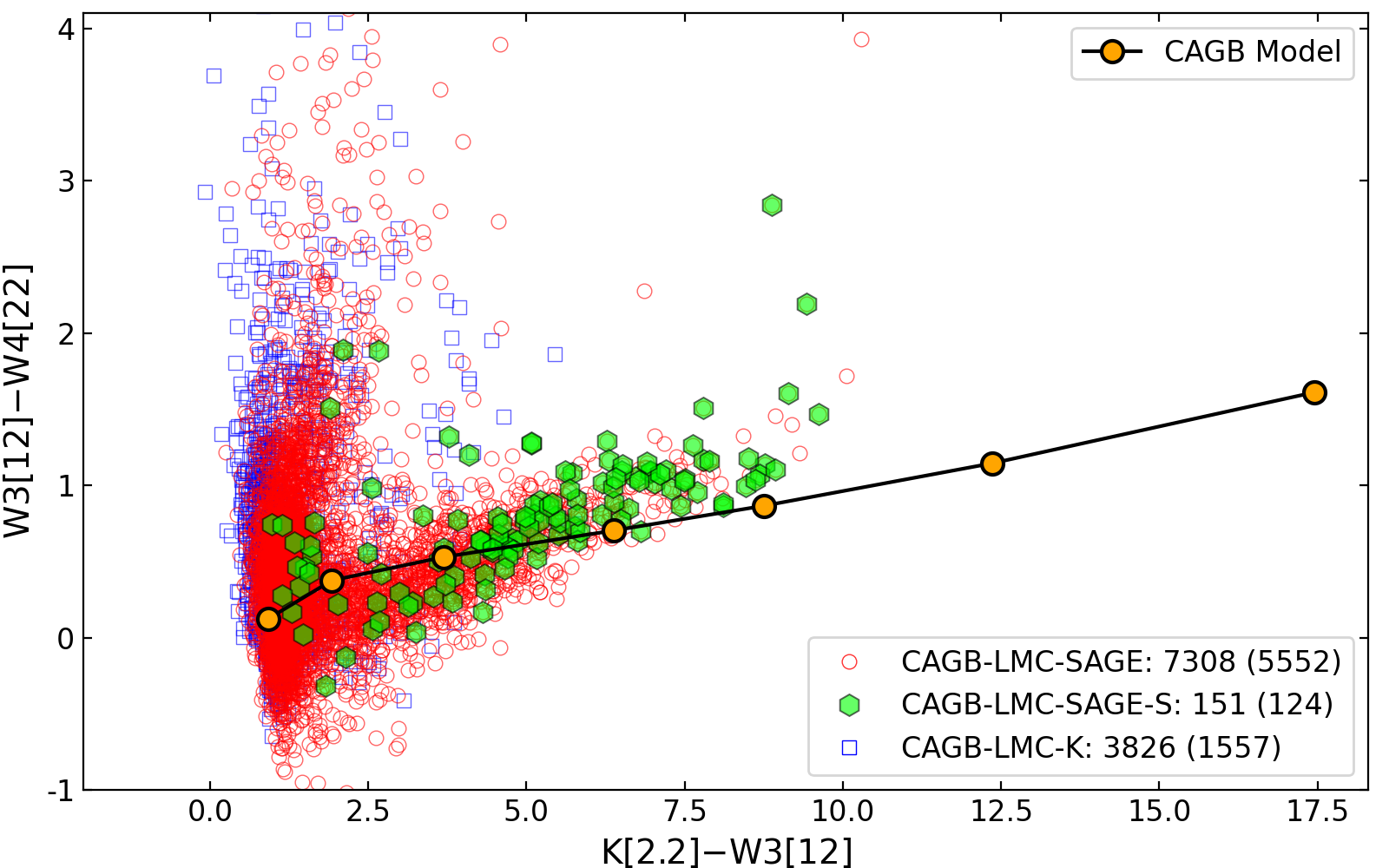}{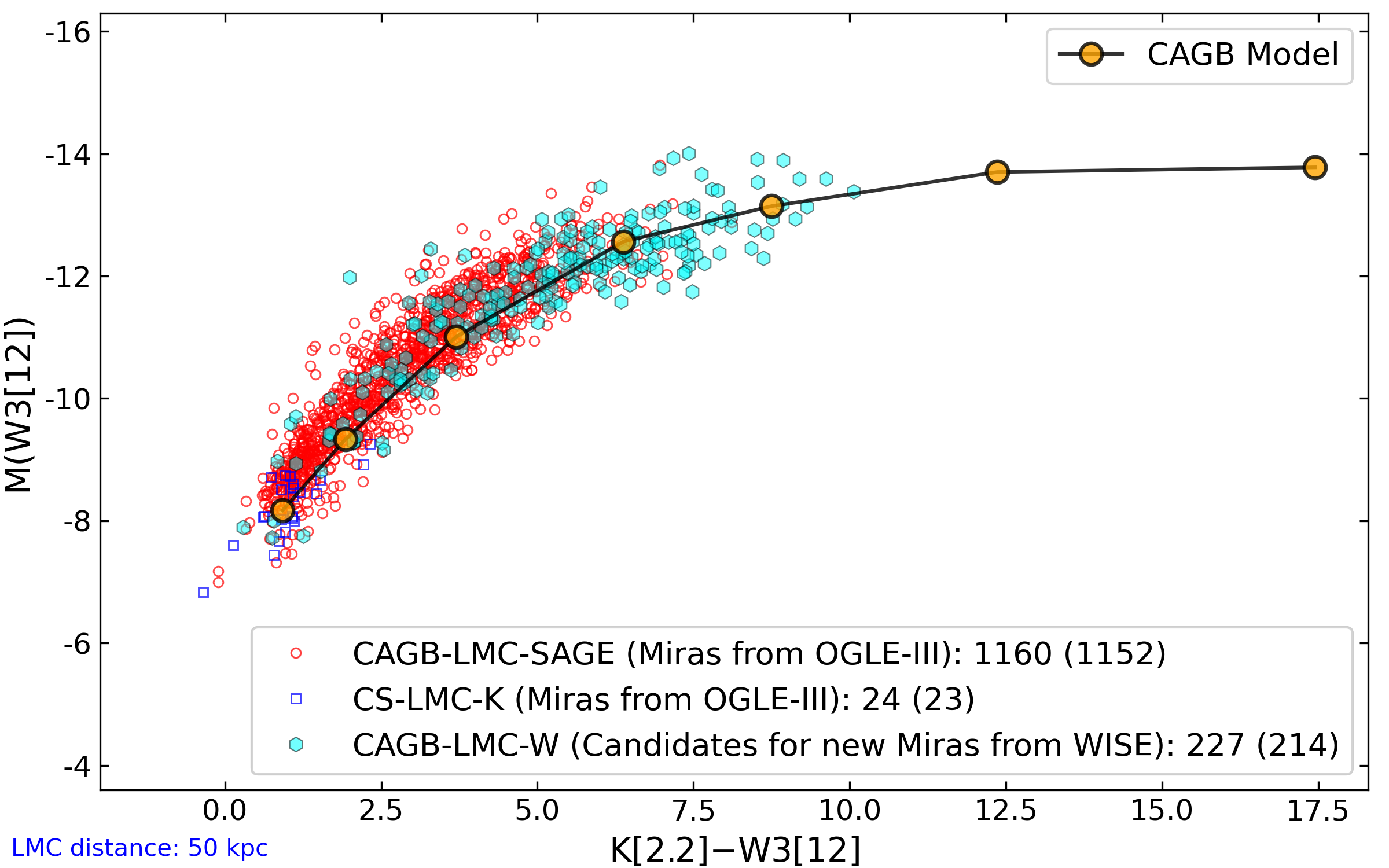}{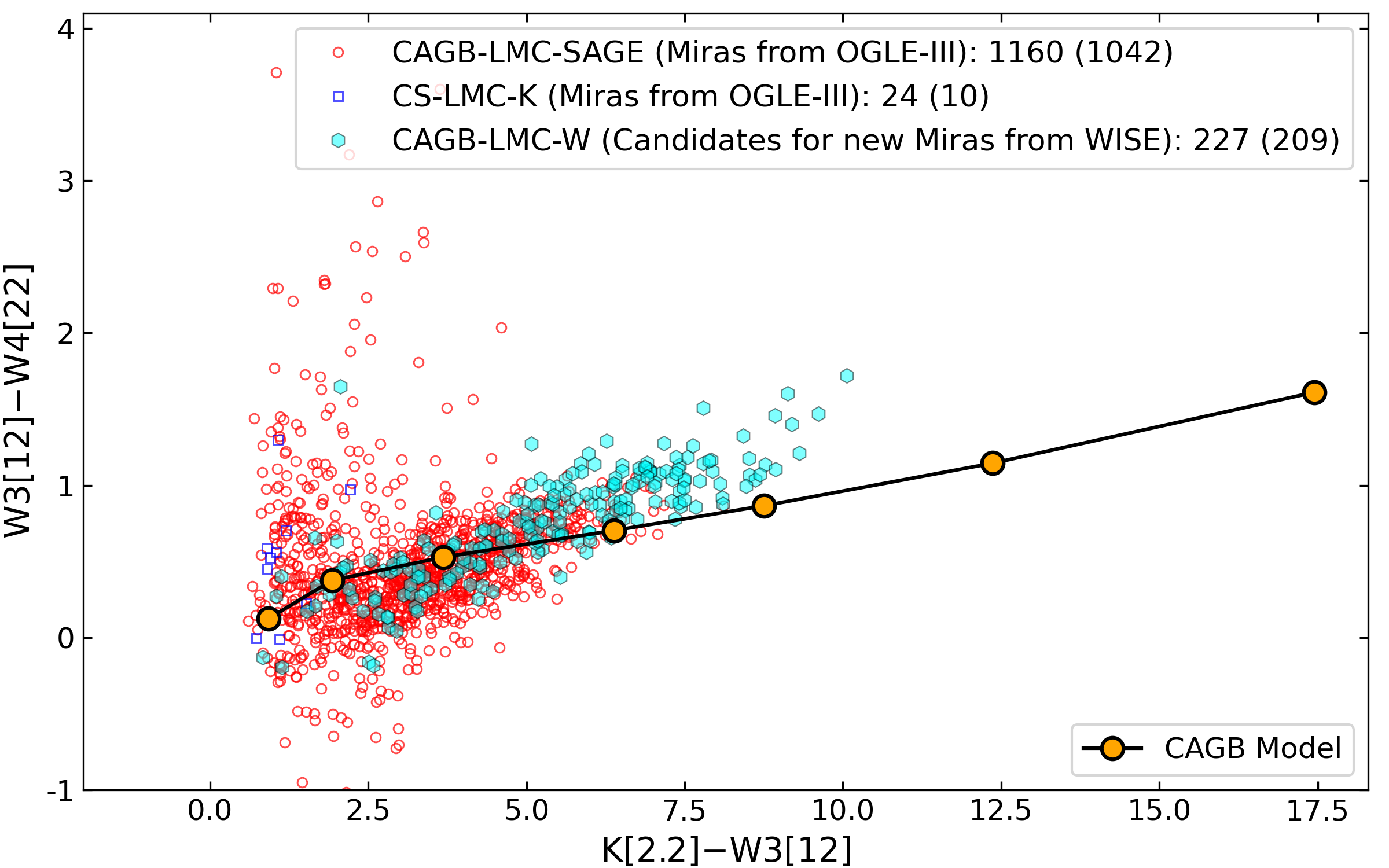}{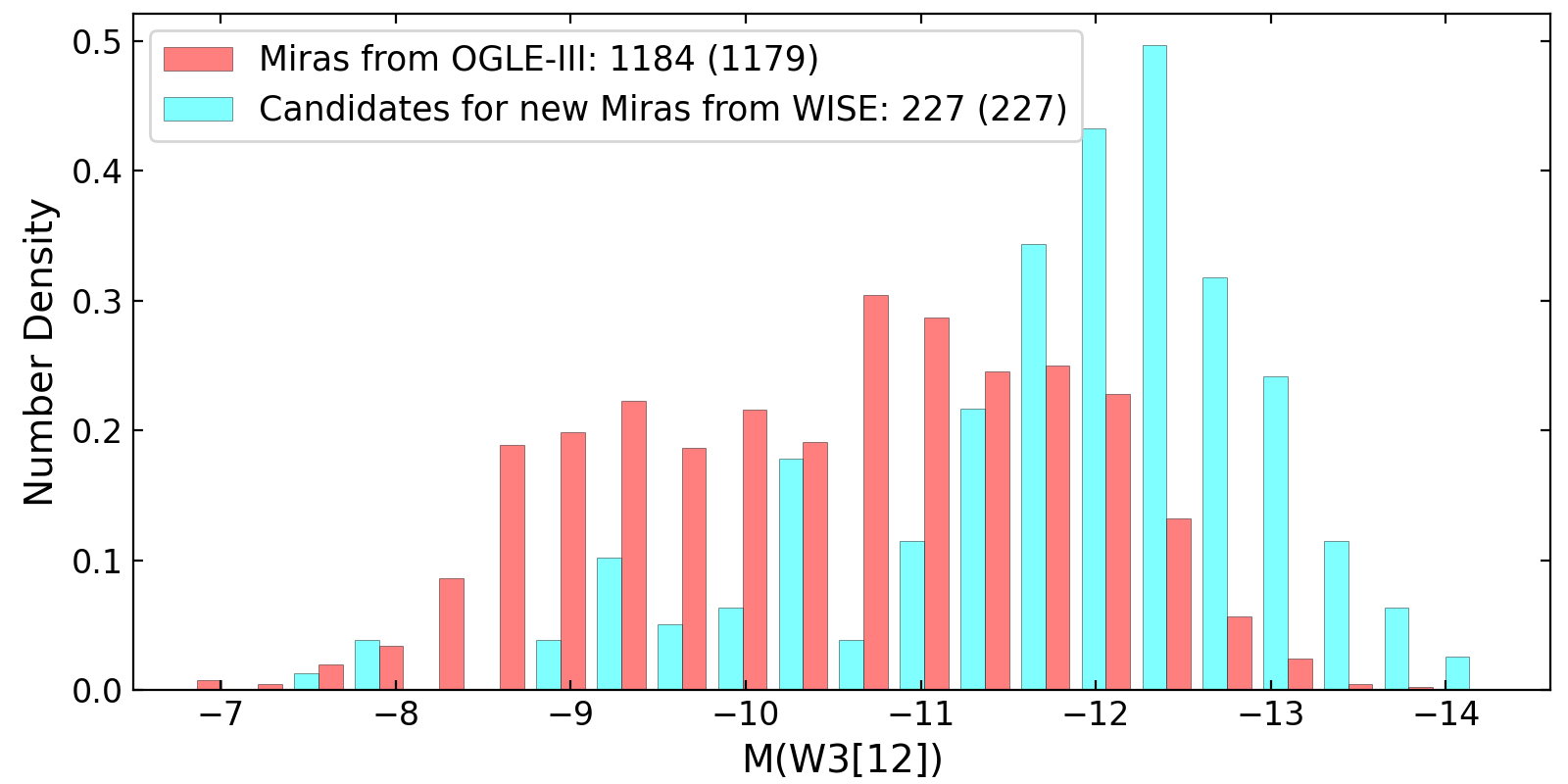}{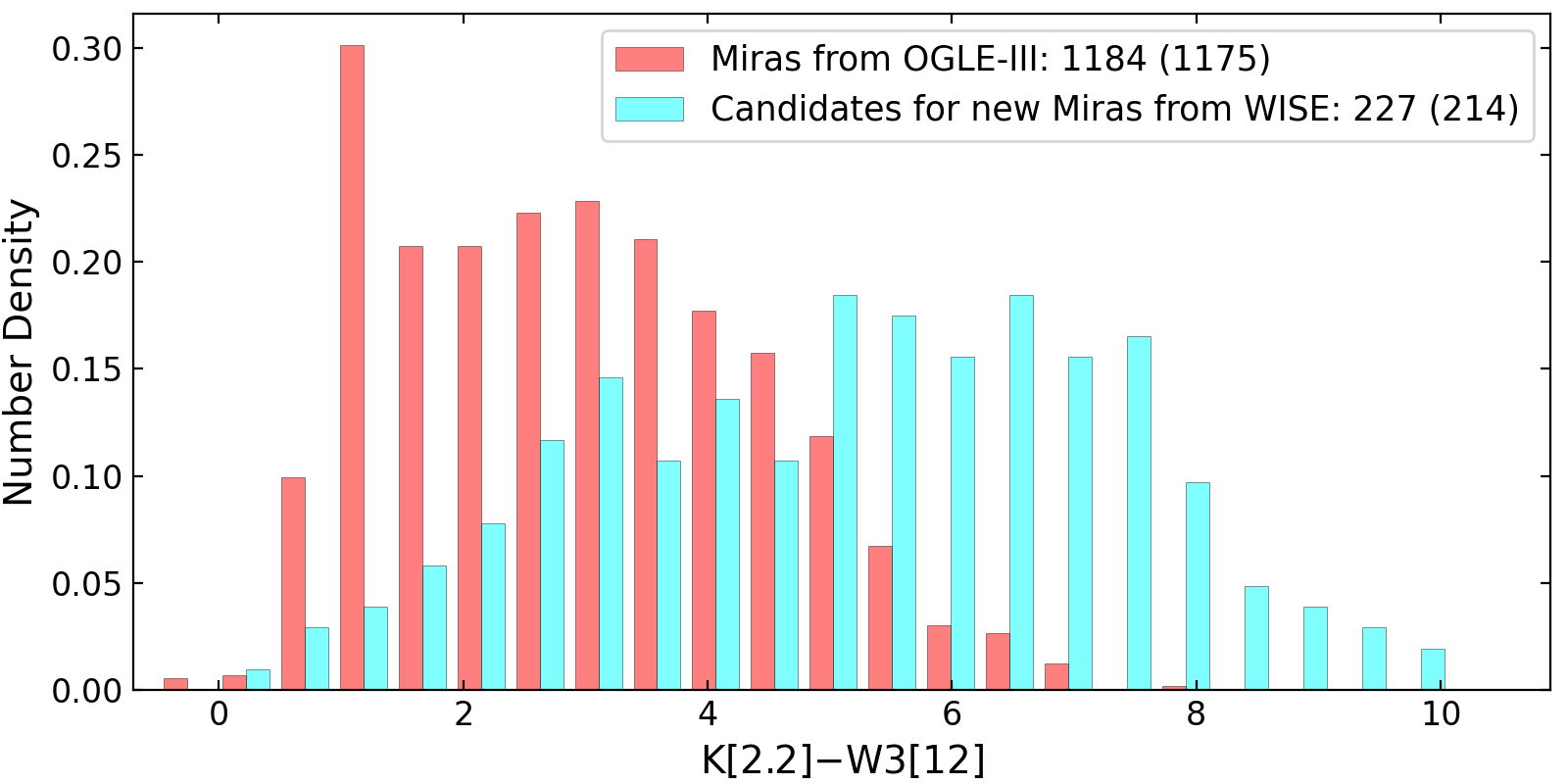}
\caption{The top panels show the IR CMD and 2CD for CAGB stars in our Galaxy.
The second row of panels presents the IR CMD and 2CD for carbon stars in the LMC.
The third row displays the IR CMD and 2CD specifically for Mira stars, selected from the CAGB stars in the LMC.
For CAGB models (AMC, $T_c$ = 1000 K): $\tau_{10}$ = 0.0001, 0.01, 0.1, 0.5, 1, 2, and 4 from left to right (see Section~\ref{sec:models}).
The bottom panels show histograms of IR magnitudes and colors for these Mira stars in the LMC.
The total number of objects for each subgroup is indicated, with the number in parentheses
representing the count of plotted objects with available observational data.
See Section~\ref{sec:cmd} for details.}
\label{f8}
\end{figure*}

\section{The IR CMD and 2CD for Carbon Stars\label{sec:cmd}}

For comparison, CAGB stars in the Milky Way can provide useful reference for 
studying carbon stars in the LMC. For CAGB stars in the Milky Way (CAGB-MW 
objects; see Table~\ref{tab:tab1}), we use the catalog of the CAGB stars by 
\citet{suh2024}.

The top panels of Figure~\ref{f8} show the CMD and 2CD for CAGB stars in the
Milky Way (CAGB-MW). They plot absolute magnitude in W3[12] and W3[12]$-$W4[24]
color against K[2.2]$-$W3[12] color. The IR-CAGB\_IC objects form a subset of the
CAGB-MW population and represent IR-bright CAGB stars with IRAS counterparts (see
\citealt{suh2024}). These stars are believed to be more evolved or massive, with
thicker dust envelopes. In the CMD, the region corresponding to extrinsic carbon
stars is marked with brown lines, following the criteria used by \citet{suh2024}.

The second row of panels shows the IR CMD and 2CD for CAGB stars in the LMC (see
Section~\ref{sec:cmd}). Compared to the Milky Way, the LMC appears to be
deficient in CAGB stars with thick dust envelopes. This may be due to less active
high-mass star formation in the LMC (e.g., \citealt{suh2020}). Some of the
CS-LMC-K objects are in the region of extrinsic carbon stars on the CMD. They are
likely to be extrinsic carbon stars (Ba or CH stars or dwarf carbon stars) which
are not in the AGB phase (see Section~\ref{sec:intro}; \citealt{suh2024}).

In the upper-left region of the IR 2CD, a group of CAGB objects in the LMC 
deviates significantly from theoretical models assuming $T_c = 1000$ K. In 
contrast, only a few CAGB stars in the Milky Way exhibit similar behavior. This 
region is likely associated with detached dust envelopes, where $T_c$ is much 
lower, around 200 to 300 K (see \citealt{suh2020}). Compared to the Milky Way, 
the LMC appears to host a larger population of CAGB stars with detached, thin 
dust envelopes.

The third row of panels in Figure~\ref{f8} presents the IR CMD and 2CD for C-rich
Mira variables in the LMC, including both OGLE-III identifications and new
candidates from WISE data. These variables align more closely with theoretical
CAGB models.

The bottom panels show histograms of IR magnitudes and colors for Mira stars in
the LMC. Miras identified from OGLE-III are based on optical observations
(I[0.8]), while the newly identified candidates are selected from WISE
observations in the IR bands (W1[3.4] and W2[4.6]). These WISE-selected
candidates generally exhibit thicker dust envelopes (or are more massive or more
evolved), and therefore tend to have redder K[2.2]$-$W3[12] colors and brighter
M(W3[12]) magnitudes.

\section{summary\label{sec:sum}}

We investigated the variability of carbon stars in the LMC. Our sample consists
of 11,134 carbon stars identified in both visual and IR bands. Among these, 1,184
objects are known Mira variables based on OGLE-III observations.

We studied the IR variability of the entire sample using WISE photometric data
spanning the past 16 years, including the AllWISE multiepoch data and the
NEOWISE-R 2024 final data release. We generated light curves using WISE
observations in the W1[3.4] and W2[4.6] bands and computed Lomb-Scargle
periodograms for all sample stars.

From the WISE light curves, variability parameters were reliably derived for
1,615 objects. Among these, 672 exhibit distinct Mira-like variability: 445
correspond to previously identified Miras from OGLE-III, while 227 are candidates
for newly identified Mira variables based on the WISE data.

We derived PMRs and PCRs in both visual and IR bands for known Mira variables as
well as newly identified candidates from WISE. The C-rich Mira variables in the
LMC exhibit strong correlations, especially in the IR bands, where the trends are
well captured by quadratic fits, though linear fits also provide reasonable
approximations. Notably, the PMR in the W3[12] band is equally well described by
both second-order and linear models. These relations provide valuable benchmarks
for investigating carbon stars in other galaxies, including the Milky Way.

We compared theoretical radiative transfer models for CAGB stars with
observational data of carbon stars in the LMC using IR CMDs and 2CDs. The IR CMD
and 2CD of C-rich Mira variables, identified by OGLE-III and complemented by new
candidates from WISE, show closer agreement with the theoretical CAGB models.


\acknowledgments I thank the anonymous reviewer for constructive comments and
suggestions. This research was supported by Basic Science Research Program
through the National Research Foundation of Korea (NRF) funded by the Ministry of
Education (RS-2022-NR075638). This research has made use of the VizieR catalogue
access tool, CDS, Strasbourg, France. This research has made use of the NASA/
IPAC Infrared Science Archive, which is operated by the Jet Propulsion
Laboratory, California Institute of Technology, under contract with the National
Aeronautics and Space Administration.

\end{document}